\documentclass[journal,draftclsnofoot,onecolumn,12pt]{IEEEtran}%

\usepackage{amsthm,amssymb,graphicx,multirow,amsmath,color,amsfonts}
\usepackage[update,prepend]{epstopdf}
\usepackage[noadjust]{cite}
\usepackage[latin1]{inputenc}
\usepackage{tikz}
\usetikzlibrary{arrows,calc}		
\usepackage{bbm} 
\usepackage{pdfpages}
\usepackage{tabulary}
\usepackage{multirow}
\usepackage{comment}

\usepackage{todonotes}

\def\nb0{{\mathbf{0}}}
\def\nb1{{\mathbf{1}}}

\def\nrmd{{\rm d}}

\def\nrmp{{\rm p}}

\def\sinc{{\rm sinc}}

\newtheorem{lemma}{Lemma}

\newtheorem{theorem}{Theorem}
\newtheorem{prop}{Proposition}
\newtheorem{cor}{Corollary}

\newtheorem{remark}{Remark}

\def\argmax{\operatorname{arg~max}}
\def\figref#1{Fig.\,\ref{#1}}%
\def\E{\mathbb{E}}

\def\P{\mathbb{P}}
\def\pc{\mathtt{P_c}}

\def\R{\mathbb{R}}

\def\T{\beta}							

\def\sir{\mathtt{SIR}}

\allowdisplaybreaks 

\usepackage{setspace}	

\begin{document}
\graphicspath{{./Figures/}}


\title{Poisson Cluster Process Based Analysis of HetNets with Correlated User and Base Station Locations}


\author{Mehrnaz Afshang and Harpreet S. Dhillon
\thanks{The authors are with Wireless@VT, Department of ECE, Virgina Tech, Blacksburg, VA, USA. Email: \{mehrnaz, hdhillon\}@vt.edu. The support of the US NSF (Grant CCF-1464293 and CNS-1617896) is gratefully acknowledged.  This paper   was presented in part at the  IEEE WCNC conference~\cite{WCNC2016_HetNet}. \hfill Last updated: \today.} 
}

\maketitle

\vspace{-1.2cm}
{
\begin{abstract}

This paper develops a new approach to the modeling and analysis of heterogeneous cellular networks (HetNets) that accurately incorporates coupling across the locations of users and base stations, which exists due to the deployment of small cell base stations (SBSs) at the places of high user density (termed {\em user hotspots} in this paper). Modeling the locations of the geographical centers of user hotspots as a homogeneous Poisson Point Process (PPP), we assume that the users and SBSs are clustered around each {\em user hotspot center} independently with two different distributions. The macrocell base station (BS) locations are modeled by an independent PPP. This model is consistent with the user and SBS configurations considered by 3GPP. 
Using this model,  we study  the performance of  a typical user in terms of coverage probability and {\em throughput} for two  association policies: i) {\em Policy 1}, under which a typical user is served by  the open-access BS that  provides maximum averaged received power, and ii) {\em Policy 2}, under which the typical user is served by the small cell tier if the maximum averaged received power from the open-access SBSs is greater than a certain {\em power threshold}; and macro tier otherwise. A key intermediate step in our analysis is the derivation of  distance distributions from a typical user to the open-access and closed-access interfering SBSs. Our analysis demonstrates that  as the number of SBSs reusing the same resource block increases, coverage probability decreases whereas  {\em throughput} increases. Therefore, contrary to the usual assumption of orthogonal channelization, it is reasonable to assign the same resource block to multiple SBSs in a given cluster as long as the coverage probability remains acceptable. This approach to HetNet modeling and analysis significantly generalizes the state-of-the-art approaches that are based on modeling the locations of BSs and users by independent PPPs. 

\end{abstract}

\begin{IEEEkeywords}
Stochastic geometry, Poisson cluster process,  user-centric capacity-driven deployment. 
\end{IEEEkeywords}
\section{Introduction} \label{sec:intro}

Current cellular networks are undergoing a significant transformation from coverage-driven deployment of macrocells to a more user-centric capacity-driven deployment of several types of low-power BSs, collectively called small cells, usually at the locations of high user density (termed {\em user hotspots} in this paper)~\cite{access2010further,3gpp2}. The resulting network architecture consisting of one or more small cell tiers overlaid on the macrocellular tier is referred to as a HetNet \cite{Chandrasekhar2008,3GPPheterogeneous2011,AndrewsSevenways2013,HetHetNets2015}. The increasing irregularity in the BS locations has led to an increased interest in the use of random spatial models along with tools from stochastic geometry and point process theory for their accurate modeling and tractable analysis; see~\cite{mukherjee2014analytical,elsawy2013stochastic, andrews2016primer, elsawy2016modeling} for detailed surveys on this topic. The most popular approach in this line of work is to model the locations of different classes of BSs as independent PPPs and perform analysis for a typical user assumed to be located independently of the BS locations. This model was first proposed in~\cite{dhillon2012modeling,5743604} for downlink analysis of HetNets, and has been extended to many scenarios of interest in the literature, see~\cite{mukherjee2014analytical,elsawy2013stochastic, andrews2016primer} and the references therein. 
Despite the success of this approach for the modeling and analysis of coverage-centric deployments and conventional single tier macrocellular network~\cite{andrews2011tractable},  this  is not   quite accurate for modeling  user-centric deployments, where the SBSs may be deployed at the user hotspots~\cite{access2010further}. In such cases, it is important to accurately capture non-uniformity as well as coupling across the locations of the users and SBSs. Developing a comprehensive framework to facilitate the analysis of such setups is the main focus of this paper.

\subsection{Related Work}  \label{sec:related}
The stochastic geometry-based modeling and analysis of HetNets has taken two main directions in the literature. The first and more popular one is to focus on the {\em coverage-centric} analysis of HetNets, in which the user locations are assumed to be independent of the BS locations~\cite{mukherjee2014analytical,elsawy2013stochastic, andrews2016primer, elsawy2016modeling,RenzoLuMIMO2015}. As noted above already, the locations of the different classes of BSs are modeled as independent PPPs. 
 This approach has been extensively used for the  analysis of key performance metrics such as coverage/outage probability~\cite{dhillon2012modeling,mukherjee2012distribution,jo2012heterogeneous,Madhusudhanan2014,JointTransmissionTanbourgi, JointTransmission2014, SakrMultipointTransmission2014,HeathKountouris2013},  rate coverage~\cite{OffloadingSingh,Rate6658810,MIMORate6881662,EnergyEfficientSoh2013}, average rate~\cite{RenzoRateHeterogeneous2013}, and network throughput~\cite{ThroughputQuek2012}. In addition, this approach enables the analytic treatment of numerous different aspects of both conventional single-tier networks as well as HetNets.
 For example, self-powered HetNets where SBSs are  powered  by a self-contained energy harvesting  module were modeled and analyzed in \cite{Harvesting6786061}. The downlink coverage and error probability analyses of  multiple input multiple output (MIMO) HetNets, where BSs are equipped with multiple antennas  were performed in~\cite{MIMORate6881662,MIMO6596082,LiZhangLetaief2014,LiZhangAndrews2016,RenzoMIMO2014}. The joint-transmission cooperation in HetNets was analyzed in~\cite{JointTransmissionTanbourgi, JointTransmission2014, SakrMultipointTransmission2014,BaccelliGiovanidisCOMP2015}. Since this line of work is fairly well-known by now, we refer the interested readers to~\cite{mukherjee2014analytical,elsawy2013stochastic, andrews2016primer, elsawy2016modeling} for more extensive surveys as well as a more pedagogical treatment of this general research direction.

 The second direction focuses on developing tractable models to study user-centric capacity-driven small cell deployments, where  SBSs are deployed at the areas of high user density. Due to the technical challenges involved in incorporating this coupling across the locations of the users and SBSs, the contributions in this direction are much sparser than above.
 One key exception is the generative model proposed  in~\cite{NonUniformDhillon}, where the BS point process is {\em conditionally thinned} in order to push the reference user closer to its serving BS, thus introducing coupling between  the locations of BSs and users. 
  While this model captures the clustering nature of users in the hotspots,  it is restricted to single-tier networks and generalization to HetNets is not straightforward. 
  Building on our recent contributions in developing analytical tools for Poisson cluster process (and Binomial point process) \cite{AfshDhi2015MehrnazD2D1,afshang2015fundamentals,AfshDhiBPP2016,AfshSahDhi2016Contact}, we recently addressed this shortcoming and generalized the analysis of non-uniform user distribution to HetNets by modeling the locations of users as a Poisson cluster process, where the correlation between user and BS locations is captured by placing BSs at the cluster centers~\cite{DownlinkChiranjib2016,SahaAfshDh2016}\footnote{
  The analytical tools developed in \cite{AfshDhi2015MehrnazD2D1,afshang2015fundamentals,AfshDhiBPP2016} are also being adopted to analyze the performance of clustered networks in the emerging paradigms in cellular communication, such as the uplink non-orthogonal multiple access~\cite{tabassum2016modeling}.}.
Although the models proposed in~\cite{NonUniformDhillon, DownlinkChiranjib2016,SahaAfshDh2016,Mankar2016,tabassum2016modeling}  accurately characterize the coupling between the user and BS locations, the assumption of modeling small cell locations with a PPP is not quite accurate in the case of user-centric deployments. This is because some user hotspots are by nature large, which necessitates the need to deploy multiple SBSs to cover that area, thus introducing {\em clustering} in their locations. Similarly, subscriber-owned SBSs, such as femtocells, are deployed on the scale of one per household/business, which naturally increases their density within residential and commercial complexes. Due to this {\em clustering},
   Poisson cluster process becomes preferred choice for modeling SBS locations in user hotspots~\cite{spatialmodelingAndrews2010,lee2013stochastic}.
 While the effect of BS clustering has been studied in \cite{MultiChannel2013,HetPCPGhrayeb2015, bacstuug2016edge,Wang2016,Heterogeneous2015,chen2013downlink,mankar2016coverage,DengHaenggiHeterogeneous2015},  none of these works provide {\em exact} analytic characterization of interference and key performance metrics for these networks. 
  More importantly, the coupling between user and BS locations, which is the key in user-centric capacity-driven deployments, has not been truly captured in these works.  For instance, the locations of users are assumed to be independent of  BS point process in \cite{HetPCPGhrayeb2015}. On the other extreme, the users are assumed to be located at a fixed distance from its serving BS in~\cite{chen2013downlink,mankar2016coverage,DengHaenggiHeterogeneous2015}.  
In this paper, we address this shortcoming by developing a new model for user-centric capacity-driven deployments that is  realistic as well as tractable. The ability of the proposed analytical model to incorporate coupling across the locations of users and BSs bridges the gap between the spatial models used by the industry (especially for user hotspots), e.g., 3GPP~\cite{access2010further}, in their simulators, and the ones used by the stochastic geometry community (see above for the detailed discussion) for the performance analysis of HetNets. As discussed  next, the main novelty is the use of Poisson cluster process for modeling both the users and the SBSs.

%
 
\subsection{Contributions and Outcomes} \label{sec:contributions}
\subsubsection*{Tractable model for user-centric capacity-driven deployment of HetNets}We develop a realistic analytic framework to study the performance
of user-centric  small cell deployments. In particular, we consider a   two-tier HetNet, comprising of a tier of small cells overlaid on
a macro tier, where macro BSs  are distributed as an independent PPP. To {capture} the coupling between the locations of SBSs and users, we model the geographical centers of user hotspots as an independent PPP around which the SBSs and users form clusters with two independent general distributions. In this setup, the candidate serving BS in each open-access tier is the one that is nearest to the typical user. From the set of candidate serving BSs,  the serving BS is chosen  based on two association  policies: i) {\em Policy 1}, where  the serving BS is the one that provides the maximum average received power to the typical user, and ii) {\em Policy 2}, where the typical user is   served by small cell tier if the maximum average received power from open-access SBSs is greater than a certain {\em power threshold}; and macro BS otherwise.


%

\subsubsection*{Coverage probability and  throughput analysis} We derive {\em exact} expressions for coverage probability of a typical user and  throughput of the whole network under
the two association policies described above. A key intermediate step in the analysis is the derivation of distance distributions from the  typical user to its serving BS,  open-access interfering SBSs, and closed-access interfering SBSs for the two association policies. Building on the tools developed for PCPs in~\cite{AfshDhi2015MehrnazD2D1}, we prove that the distances from open-access interfering SBSs conditioned on the location of the serving BS and the typical user are independently and identically distributed (i.i.d.).  
Using this i.i.d. property, the Laplace transform of  interference distribution is obtained, which then enables the derivation of the coverage probability and  throughput results.
\subsubsection*{System design insights} Our analysis leads to several useful design insights. 
First, it reveals that more aggressive frequency reuse within a given cluster (more SBSs reusing the same resource blocks in a given cluster) has a conflicting effect on the  coverage and throughput: 
 throughput increases and coverage decreases. This observation shows that more SBSs can  reuse the same resource blocks in a given cluster as long as the coverage probability is acceptable.
Thus, the strictly orthogonal  resource allocation strategy  that allocates each resource block to at most one SBS in a given cluster (e.g., see~\cite{HetPCPGhrayeb2015}) may not be  efficient in terms of throughput for this setup.
Second,  our analysis reveals that there exists an optimal {\em power threshold} that maximizes the coverage probability of a typical user under association Policy 2. 

\section{System Model} \label{sec:SysMod}
We consider a two-tier heterogeneous cellular network  consisting of macrocell and small cell BSs, where SBSs and users are clustered around geographical centers of user hotspots, as shown in \figref{Fig: system model Figs2}(c). This  model  is inspired by the fact that several SBSs may be required to be deployed  in each {user hotspot (hereafter referred to as cluster)}  in order to handle mobile data traffic generated in that user hotspot~\cite{access2010further}. The analysis is performed for a typical user, which is the  user chosen uniformly at random from amongst all users in the network. Throughout this paper, the cluster in which the typical user is located will be referred to as the {\em representative cluster.}
For this setup, we assume that the typical user is allowed to connect to any macro BS in the whole network and any SBS located within the representative cluster. Other SBSs (the ones located outside the representative cluster) simply act as interferers for the typical user. This setup is inspired by the situations in which SBSs are enterprise owned BSs intended to serve only the authorized users (who have permission to connect to that network). Therefore, we will refer to the macro BSs and the SBSs within the representative cluster as {\em open access} BSs (with respect to the typical user) and the rest of the SBSs as {\em closed access} BSs.
%
 Note that while our model is, in principle, extendible to completely open access $K$-tier heterogeneous cellular networks, we limit our discussion to this two-tier setup for the simplicity of both notation and exposition.

\begin{figure}       
   \centering{
    \includegraphics[width=.3\textwidth]{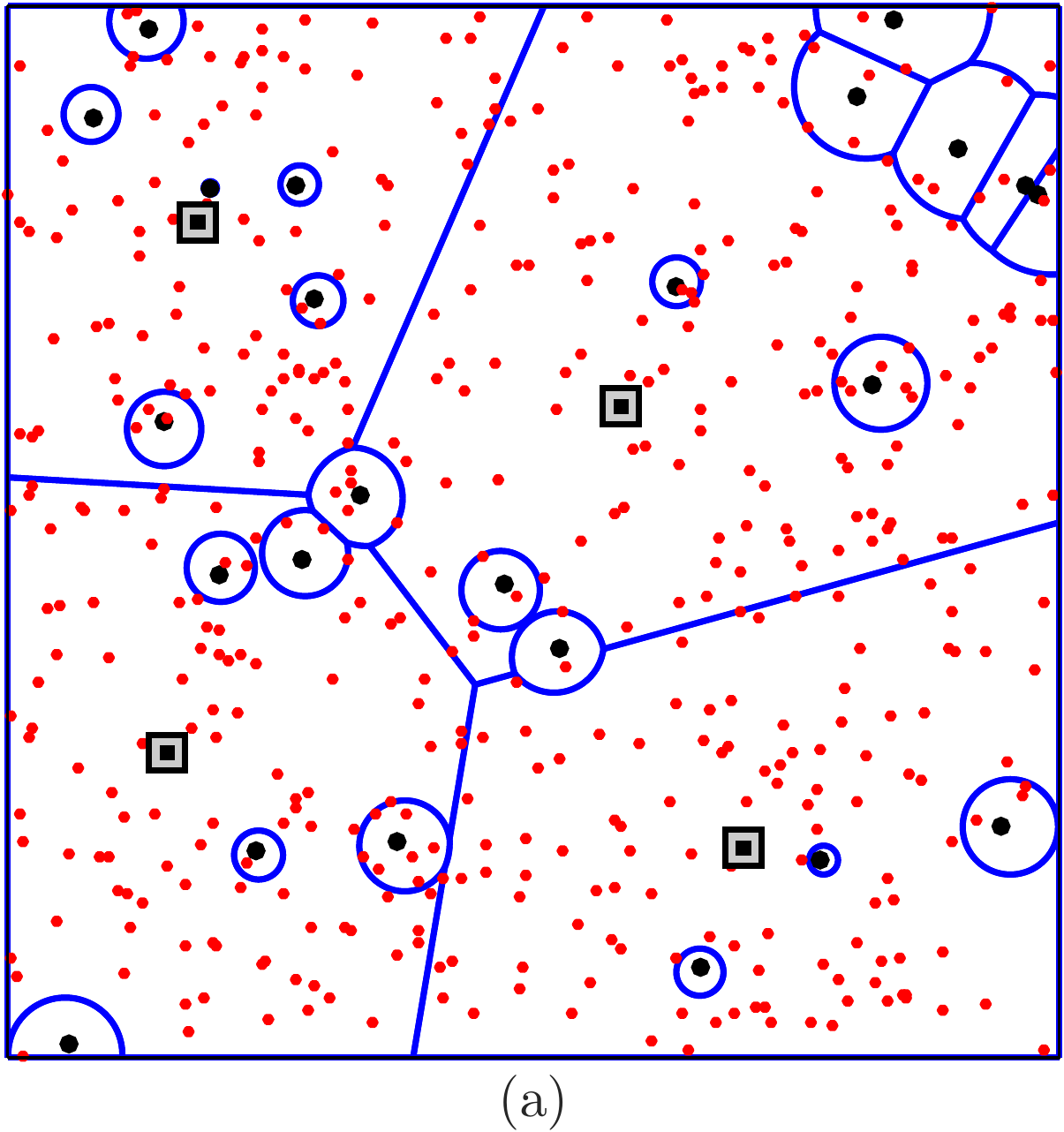}
  \includegraphics[width=.3\textwidth]{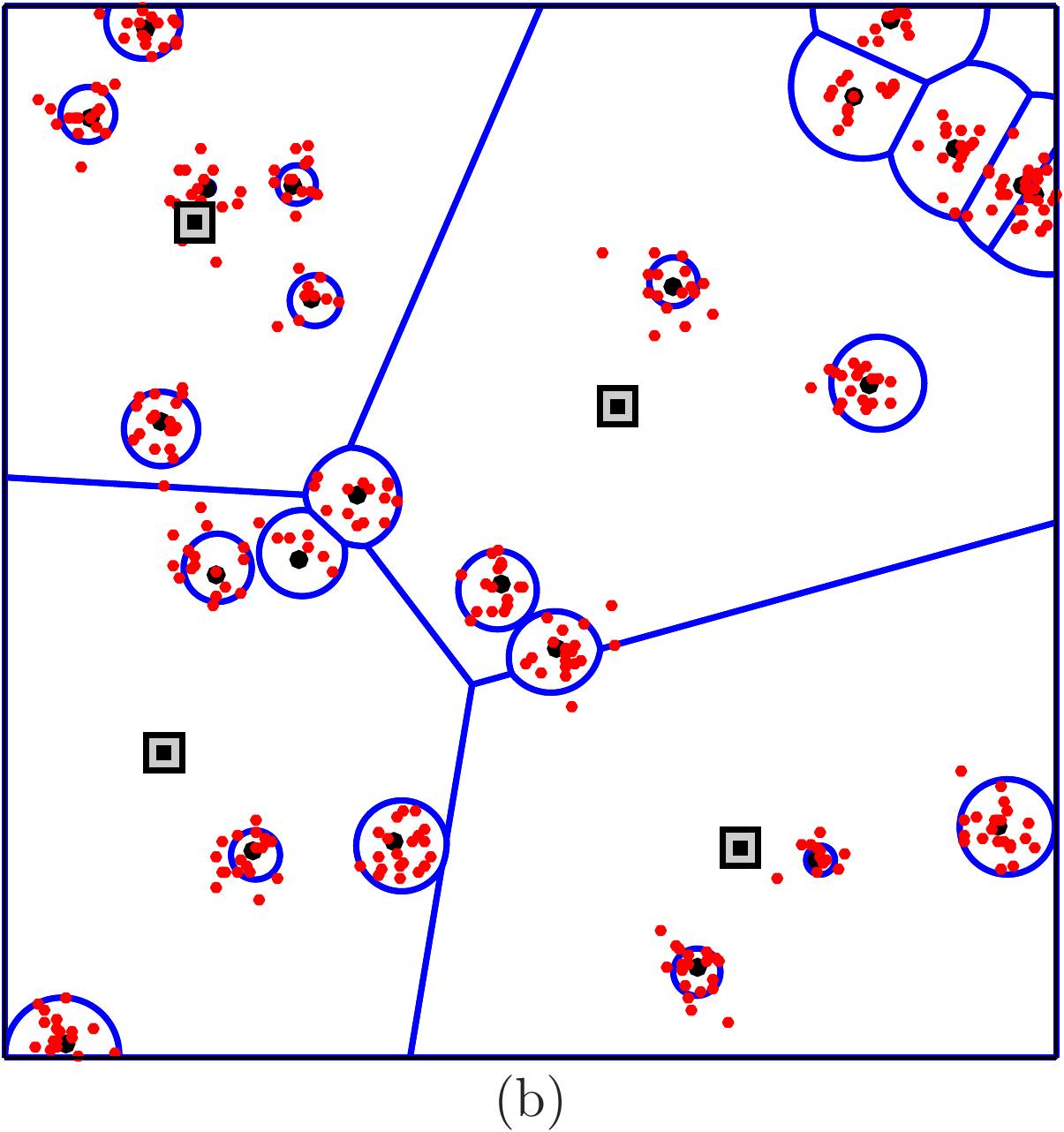}
  \includegraphics[width=.3\textwidth]{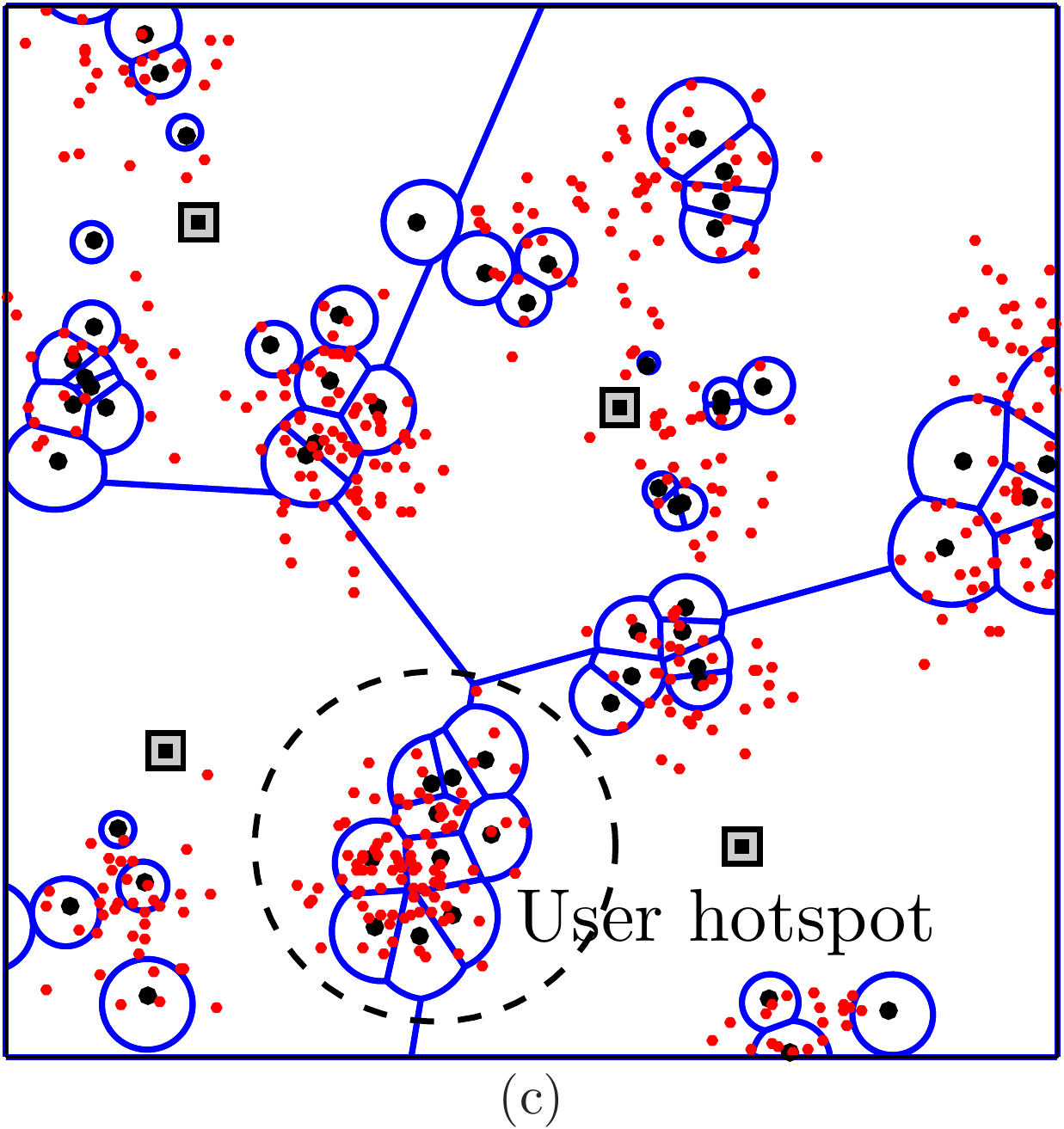}
              \caption{ (a) Coverage-centric deployments where users are uniformly and independently distributed (our original $K$-tier HetNet model~\cite{dhillon2012modeling}), (b) user-centric deployment where users are clustered around SBSs (recent enhancement to the $K$-tier model~\cite{SahaAfshDh2016}), and (c) user-centric deployment where users and SBSs are  clustered around  geographical centers of user hotspots (this paper).   Black squares, black dots, and small red dots denote the locations of macro BSs, SBSs, and users, respectively.}
                   \label{Fig: system model Figs2} 
                   }   
 \end{figure}

\subsection{Spatial Setup and Key Assumptions}
We model  the locations of  macro  BSs as an independent homogeneous PPP $\{{\bf z}_{\rm m}\} \equiv \Phi_{\rm m}$ with density $\lambda_{\rm m}$. In order to capture  the coupling between the locations of SBSs and users in hotspots, we model the locations of SBSs and users 
as  two  Poisson cluster processes with  the same parent  point process, where the later models the geographical centers of user hotspots.  
 It should be noted that in reality user distribution is a superposition of homogeneous and non-homogeneous distributions. For instance,  pedestrians and users in transit are  more likely to be uniformly distributed in the network  and hence  homogeneous PPP is perhaps a better choice for the analysis of such users.    On the other hand, users in hotspots  exhibit clustering behavior for which Poisson cluster process is a more appropriate model than a homogeneous PPP~\cite{SahaAfshDh2016}.
  The framework provided in this paper can be extended to the case of mixed user distribution consisting of both homogeneous and non-homogeneous user distributions without much effort.   Besides, the analysis of homogeneous user distributions in such setups is well known \cite{dhillon2012modeling,mukherjee2014analytical,elsawy2013stochastic, andrews2016primer}, which is the reason we chose to focus on the more challenging case of non-homogeneous user distributions in which
the user and SBS locations are coupled.

 
 Poisson cluster process can be formally defined as a union of  offspring points which are  independent of each other, and identically distributed  around parent points~\cite{DalVerB2003,ganti2009interference}. Modeling the locations of  parent point process (i.e., cluster centers)  as  a homogeneous  PPP $\{{\bf x}\} \equiv \Psi_{\rm p}$  with density $\lambda_{\rm p}$,

\begin{enumerate}
\item  the set of users within  a cluster centered at ${\bf x} \in \Psi_{\rm p}$  is denoted by $\{{\bf y}_{\rm u}\}\equiv {\cal N}^{\bf x}_{\rm u}$ (with ${\bf y}_{\rm u} \in \R^2$),  where each set contains  a sequence of i.i.d. elements conditional on $\mathbf{x}$ (denoting locations), and the PDF of each element is $f_{{\bf Y}_{\rm u}}({\bf y}_{\rm u})$, and
\item  the set of SBSs within  a cluster centered at ${\bf x} \in \Psi_{\rm p}$  is denoted by $\{{\bf y}_{\rm s}\}\equiv {\cal N}^{\bf x}_{\rm s}$ (with ${\bf y}_{\rm s} \in \R^2$),  where each set contains  a sequence of i.i.d. elements conditional on $\mathbf{x}$, and the PDF of each element is $f_{{\bf Y}_{\rm s}}({\bf y}_{\rm s})$.
\end{enumerate}
The locations of SBSs ${\cal N}^{\bf x}_{\rm s}$ and users ${\cal N}^{\bf x}_{\rm u}$ conditioned on  ${\bf x} \in \Psi_{\rm p}$ are independent. For this setup,
after characterizing all theoretical results in terms of general distributions $f_{{\bf Y}_{\rm s}}({\bf y}_{\rm s})$ and $f_{{\bf Y}_{\rm u}}({\bf y}_{\rm u})$, we specialize the results to Thomas cluster process~\cite{haenggi2012stochastic} in which the points are distributed around cluster centers according to an independent Gaussian distribution: 
\begin{align}\label{eq: Thomas location PDF}
   f_{{\bf Y}_{ \ell}}({\bf y}_{ \ell})= \frac{1}{2 \pi \sigma_{ \ell}^2}\exp \left(-\frac{\|{\bf y}_{\ell}\|^2}{2 \sigma_{ \ell}^2}\right),  \:\text{where}\: {\ell}\in\{{\rm s},{\rm u}\}.
\end{align}
From  the set of SBSs located in the cluster centered at ${\bf x} \in \Psi_{\rm p}$, we assume that the subset of ${\cal B}^{\bf x}_{\rm s} \subseteq	 {{\cal N}^{\bf x}_{\rm s}}$ reuse the same resource block.
This subset will  be henceforth referred to as  a {\em set of simultaneously active SBSs},  where the number of simultaneously active SBSs $|{\cal B}^{\bf x}_{\rm s}|$ is assumed to have a Poisson distribution with mean $\bar{n}_{\rm as}$. 
Denote by  ${{\bf x}_0}\in \Psi_{\rm p}$ the location of the center of representative cluster. In order to simplify the {\em order statistics} arguments that will be used in the selection of candidate serving BSs in the cluster located at ${\bf x} \in \Psi_{\rm p}$, we assume that  the  total number of  SBSs  (i.e., $|{{\cal N}^{{\bf x}_0}_{\rm s}}|$) in the representative cluster is fixed and equal to $n_{{\rm s}_0}$, where ${\cal B}^{{\bf x}_0}_{\rm s}\subseteq	 {\cal N}^{{\bf x}_0}_{\rm s}$ represents the set of simultaneously active SBSs in the representative cluster. Note that $|{\cal B}_{\rm s}^{{\bf x}_0}|$ is truncated Poisson random variable with maximum value being 
$n_{{\rm s}_0}$, and the serving SBS will be chosen from amongst SBS in ${\cal N}_{\rm s}^{{\bf x}_0}$. The pictorial representation of our setup along with the system  models used in the prior work are presented in \figref{Fig: system model Figs2}.

%

\subsection{Propagation Model}
We assume that all links to the typical user suffer from a standard power-law path-loss with exponent $\alpha>2$, and Rayleigh fading. Thus the received power at the typical user (located at the origin) from the $j^{th}$ tier BS (where $j\in\{{\rm s},{\rm m}\}$) located at ${\bf z}_j$  is:
\begin{align}
P_{\rm r}=P_j h_j  \|{\bf z}_j\|^{-\alpha}; \quad j\in\{{\rm s},{\rm m}\},
\end{align}
where  $\|\cdot\|^{-\alpha}$ models power-law path-loss,  $h_j$ is exponential random variable with unit mean independent of all other random variables, and $P_j$ is transmit  power,  which is assumed to be constant for the BSs in tier $j\in \{{\rm s},{\rm m}\}$. Note that  index `${\rm m}$' and `${\rm s}$' refer to  macro tier and small cell tier, respectively. {Denote  by   $\{{\bf z}_{\rm s}={\bf x}_0+{{\bf y}_{\rm s}}; {{\bf y}_{\rm s}} \in {\cal N}^{{\bf x}_0}_{\rm s} \} \equiv \Phi_{\rm s}$  the locations of open-access SBSs.} The candidate serving BS location  from $\Phi_j$ is:}
\begin{align}
{{\bf z}^{*}_j}= \argmax_{{{\bf z}_j}\in \Phi_j} P_j \|{\bf z}_j\|^{-\alpha},
\end{align}
where ${{\bf z}^{*}_j}$ is the location of the nearest  open-access BS  of the $j^{th}$ tier (i.e.,  $\Phi_j$) to the typical user. 
In order to select the  serving BS from amongst the set of candidate serving BSs, we consider two association policies. In Policy 1, the goal is to maximize coverage probability  and hence the reference signal received power (RSRP), which is the average received power of all  open-access BSs measured by a typical user, are compered and the user is served by the BSs which provides maximum average received power.  In Policy 2, the goal is to balance the load across the network, and hence user is served by small cell tier if maximum RSRP from open access SBSs is greater than specific power threshold; and macro tier otherwise. More details on these two  association policies will be provided in the next Section.
%
 \subsubsection{$\sir$ at a typical user served by macrocell}  
Assuming that the typical user is served by the macro BS located at ${\bf z}^*$,   the total
interference seen at the typical  user originates from three sources: (i) interference caused by macro BSs (except the
serving BS) defined as: $ {\cal I}_{\rm mm}= \sum_{{\bf z}_{\rm m}\in \Phi_{\rm m}\setminus {{\bf z}^{*}}} P_{\rm m} h_{\rm m} \|{\bf z}_{\rm m}\|^{- \alpha}
$, (ii) intra-cluster interference  caused by simultaneously active open-access SBSs  inside the
representative cluster (i.e., typical user's cluster), which is defined as: $ {\cal I}_{\rm sm}^{\rm intra}= \sum_{{\bf y}_{\rm s}\in {\cal B}_{\rm s}^{{\bf x}_0}} P_{\rm s}  h_{\rm s} \|{\bf x}_0+{\bf y}_{\rm s}\|^{- \alpha}$, and (iii) inter-cluster interference caused by simultaneously active closed-access SBSs  outside the
representative cluster defined as: 
$ {\cal I}_{\rm sm}^{\rm inter}=\sum_{{\bf x} \in \Psi_{\rm p} \setminus {\bf x}_0} \sum_{{\bf y}_{\rm s}\in {\cal B}_{\rm s}^{{\bf x}}}  \\ P_{\rm s} h_{\rm s} \|{\bf x}+{\bf y}_{\rm s}\|^{- \alpha}$. The $\sir$ at the typical user conditioned on the serving BS being macrocell is:
 \begin{align}
 {\tt SIR}_{\rm m}= \frac{P_{\rm m} h_{\rm m} \|{\bf z}^{*}\|^{-\alpha}}{ {\cal I}_{\rm mm}+{\cal I}_{\rm sm}^{\rm intra}+{\cal I}_{\rm sm}^{\rm inter}}.
\end{align}
\subsubsection{$\sir$ at a typical user served by small cell}
Assuming that a typical user is served by the SBS located at  ${\bf z}^*={\bf x}_0+{\bf  y}_0$,  the contribution of the total
interference seen at the typical user can be partitioned into three sources: (i) interference from macro BSs defined as: $ {\cal I}_{\rm ms}= \sum_{{\bf z}_{\rm m}\in \Phi_{\rm m}} P_{\rm m} h_{\rm m} \|{\bf z}_{\rm m}\|^{- \alpha}
$, (ii)  interference from simultaneously active open-access SBSs (except the
serving BS) inside the representative cluster defined as: ${ {\cal I}_{\rm ss}^{\rm intra}= \sum_{{\bf y}_{\rm s}\in {\cal B}_{\rm s}^{{\bf x}_0}\setminus {{\bf y}_0}} P_{\rm s}  h_{\rm s} \|{\bf x}_0+{\bf y}_{\rm s}\|^{- \alpha}}$,  and (iii) interference from  simultaneously active closed-access SBSs outside the representative cluster defined as: $ {\cal I}_{\rm ss}^{\rm inter}=\sum_{{\bf x} \in {\Psi_{\rm p}} \setminus {\bf x}_0} \sum_{{\bf y}_{\rm s}\in {\cal B}_{\rm s}^{{\bf x}}} P_{\rm s}  h_{\rm s} \|{\bf x}+{\bf y}_{\rm s}\|^{- \alpha}$. Therefore,  the $\sir$ at the typical  user served by the small cell is:
 \begin{align}
 {\tt SIR}_{\rm s}= \frac{P_{\rm s} h_{\rm s} \|{\bf z}^*\|^{-\alpha}}{ {\cal I}_{\rm ms}+{\cal I}_{\rm ss}^{\rm intra}+{\cal I}_{\rm ss}^{\rm inter}}.
\end{align}


\begin{table}
\centering{
\caption{Summary of notation}
\scalebox{.7}{%
\begin{tabular}{c|c}
  \hline
   \hline
  \textbf{Notation} & \textbf{Description}  \\
     \hline
  $\Phi_{\rm m};\  \lambda_{\rm m}$ & Independent PPP modeling the locations of macro BSs; density of $\Phi_{\rm m}$\\
  \hline
  $\Psi_{\rm p}; \ \lambda_{\rm p}$ & Independent PPP modeling the locations of parent points (cluster centers); density of $\Psi_{\rm p}$\\
  \hline
 $ {\cal N}^{{\bf x}}_{\rm s}; \  {\cal N}^{{\bf x}}_{\rm u}$ & Set of  SBSs in a cluster centered at ${\bf x}\in \Psi_{\rm p}$; Set of  users in a cluster centered at ${\bf x}\in \Psi_{\rm p}$\\
  \hline
 ${\cal B}^{{\bf x}}_{\rm s}\subseteq	 {\cal N}^{{\bf x}}_{\rm s}; \bar{n}_{\rm as}$ & Set of  simultaneously active SBSs in a cluster centered at ${\bf x}\in \Psi_{\rm p}$  with mean $\bar{n}_{\rm as}$ \\
 \hline
  $\sigma_{\rm s}^2$ ($\sigma_{\rm u}^2$) & Scattering variance of the SBS (user) locations around each cluster center\\
  \hline
      $P_j$; $h_j$; $\alpha$; $\beta$  &Transmit power; channel power gain under Rayleigh fading;  path loss exponent; target $\sir$\\
    \hline
 
   $ {\cal A}_j^{\rm P1}$ $( {\cal A}_j^{\rm P2})$ &Association probability under association Policy 1 (Policy 2), where $j \in \{{\rm s}, {\rm m}\}$\\
    \hline
     $  {\tt P}^{\rm P1}_{{\rm c}_j}$ $(  {\tt P}^{\rm P2}_{{\rm c}_{ j}})$ & Coverage probability of a typical user served by tier $j\in\{{\rm s},{\rm m}\}$ under  association Policy 1 (Policy 2)\\
    \hline
  $  {\tt P}^{\rm P1}_{{\rm c}_{\rm T}}$ $(  {\tt P}^{\rm P2}_{{\rm c}_{ \rm T}})$ & Total coverage probability under association Policy 1 (Policy 2)\\
    \hline
  $  {\cal T}^{\rm P1}$ $(  {\cal T}^{\rm P2})$ & Throughput under  association Policy 1 (Policy 2)\\
      
   \hline
    \hline

\end{tabular}\label{table:notation-cluster HetNET}
}
} 
\end{table}


\section{Serving and Interfering Distances} \label{sec: AssociationServingInterfering}
This is the first main technical section of the paper, where we derive the association probability of a typical user to macro BSs and SBSs. We  then characterize the distributions of distances from serving and interfering {macro BSs and SBSs} to a typical user. 
These distance distributions will be used to characterize the coverage probability of a typical user, and  throughput of the whole network in the next section.  We now begin by providing relevant distance distributions.
\subsection {Relevant distance distributions}
Let us denote the distances from a typical user to its  nearest open-access SBS and macro BS by $R_{ \rm s}$ and ${R_{\rm m}}$, respectively. 
 In order to calculate the association probability and the serving distance distribution,  it is important  to first characterize the density functions of  $R_{\rm m}$ and $R_{\rm s}$. The PDF and CDF of the distance from a typical user to its nearest macro BS, i.e., $R_{\rm m}$, can be easily obtained by  using  null probability of a homogeneous PPP as~\cite{haenggi2005distances}:
\begin{align}\label{Eq: PDF macro closest}
&\text {PDF:} \quad f_{R_{\rm m}}(r_{\rm m})= 2 \pi \lambda_{\rm m} r_{\rm m} \exp(- \pi \lambda_{\rm m} r_{\rm m}^2) \quad   
&\text{ CDF:} \quad F_{R_{\rm m}}(r_{\rm m})=1- \exp(- \pi \lambda_{\rm m} r_{\rm m}^2).
\end{align}
However, characterizing  the density function of distance from a typical user to its nearest open-access  SBS, i.e., the nearest  SBS to the typical user from representative cluster,  is  more challenging.  To derive the density function of $R_{\rm s}$, it is useful to define the  sequence of distances from the typical user to the SBSs located within the representative cluster  as ${\cal D }^{{\bf x}_0}_{\rm s}=\{{u: u=\|{\bf x}_{0}+{\bf y}_{\rm s}\|}, \forall {\bf y}_{\rm s} \in {\cal N}^{{\bf x}_0}_{\rm s} \}$.  Note that the elements  in  ${\cal D }^{{\bf x}_0}_{\rm s}$ are correlated due to the common factor ${\bf x}_0$. But  this correlation can be handled by conditioning on the location of representative cluster center ${\bf x}_0$  because the SBS locations are i.i.d. around the cluster center by assumption. The conditional PDF  of any (arbitrary) element  in the set ${\cal D }^{{\bf x}_0}_{\rm s}$ is characterized next. 
\begin{lemma}  \label{lem: PDFs of intra-cluster distances}The  distances in the set  ${\cal D }^{{\bf x}_0}_{\rm s}$ conditioned on the distance of the typical user to the cluster center, i.e., $V_0=\|{\bf x}_0\|$, are i.i.d., where the CDF of each element for a given $V_0=\nu_0$ is:
\begin{align}\label{eq: CDF of U}
F_{U}(u|\nu_0)=\int_{z_1=-u}^{z_1=u}  \int_{z_2=-\sqrt{u^2-z_1^2}}^{z_2=\sqrt{u^2-z_1^2}} f_{{\bf Y}_{\rm s}}(z_1-\nu_0,z_2)   {\rm d} {z_2} {\rm d}z_1, 
\end{align}
and the conditional PDF of $U$ is:
\begin{align} \label{eq: PDF  of U}
f_{U}(u|\nu_0)=\int_{-u}^{u}   \frac{u}{\sqrt{u^2-z_1^2}} \Big[f_{{\bf Y}_{\rm s}}  \Big(z_1-\nu_0, \sqrt{u^2-z_1^2}\Big)+ f_{{\bf Y}_{\rm s}}  \Big(z_1-\nu_0, -\sqrt{u^2-z_1^2}\Big) \Big] {\rm d} z_1,
\end{align}
where the PDF of $V_0$ is given by
\begin{align} \label{eq: PDF of V_0}
f_{V_0}(\nu_0)=\int_{-\nu_0}^{\nu_0} \frac{\nu_0}{\sqrt{\nu_0^2-x_1^2}}  \Big[f_{{\bf Y}_{\rm u}} \Big(x_1, \sqrt{v_0^2-x_1^2} \Big)+f_{{\bf Y}_{\rm u}} \Big(x_1, -\sqrt{v_0^2-x_1^2} \Big)\Big] {\rm d} x_1.
\end{align}
\end{lemma}
\begin{IEEEproof}
See Appendix \ref{proof: PDFs of intra-cluster distances}.
\end{IEEEproof}
The density functions of distances presented in Lemma~\ref{lem: PDFs of intra-cluster distances} are specialized to the case of Thomas cluster process in the next Corollary.
\begin{cor}\label{cor: PDFs  intra Thomas}
 For the special case of Thomas cluster process, the distances in the set  ${\cal D }^{{\bf x}_0}_{\rm s}$ are conditionally i.i.d., with CDF
\begin{align}\label{eq: CDF of U, Rician}
 F_{U}(u|\nu_0)=1-Q_1\Big(\frac{\nu_0}{\sigma_{\rm s}}, \frac{u}{\sigma_{\rm s}}\Big), \quad u>0,
 \end{align}
 where $Q_1(a,b)$ is the Marcum Q-function defined as $Q_1(a,b) = \int_b^{\infty} t e^{-\frac{t^2 + a^2}{2}} I_0(at) {\rm d} t$, and  the  PDF of each element  is:
\begin{align}\label{eq: PDF of U, Rician}
 f_{U}(u|\nu_0)=\frac{u}{ \sigma_{\rm s}^2} \exp\left(-\frac{u^2+\nu_0^2}{2 \sigma_{\rm s}^2}\right) I_0\left(\frac{u \nu_0}{\sigma_{\rm s}^2}\right), \quad u>0,
\end{align}
where  $I_0(\cdot)$ is  the modified Bessel function of the first kind with order zero. The PDF of $V_0$ is:
\begin{align}\label{eq: PDF of V_0, Rayleigh}
f_{V_0}(\nu_0)=\frac{\nu_0}{  \sigma_{\rm u} ^2}\exp\left(-\frac{\nu_0^2}{2 \sigma_{\rm u}^2}\right), \quad \nu_0>0.
\end{align}
\end{cor}
\begin{IEEEproof}
For the special case of Thomas cluster process, the PDF of ${\bf Y}_{\rm s}$  with realization ${\bf y}_{\rm s}=(y_{\rm s_1}, y_{\rm s_2})$ can be expressed as: 
\begin{align*}
f_{{\bf Y}_s}(y_{\rm s_1}, y_{\rm s_2})=\frac{1}{2 \pi \sigma_{\rm s }^2}\exp \left(-\frac{{y_{\rm s_1}^2+ y_{\rm s_2}^2}  }{2 \sigma_{ \rm s}^2}\right).
\end{align*}
Substituting $f_{{\bf Y}_s}(\cdot)$ into \eqref{eq: PDF  of U} and letting $z_1=u \cos{\theta}$, we get Rician distribution as:
\begin{align*}
f_{U}(u|\nu_0)= \frac{u}{ \sigma_{\rm s }^2} \exp(-\frac{u^2+\nu_0^2 }{2 \sigma_{\rm s }^2}) \underbrace{\int_{-\pi/2}^{\pi_2} \frac{1}{2 \pi } \exp(\frac{ u \nu_0 \cos \theta}{ \sigma_{\rm s }^2})}_{I_0(\frac{u \nu_0}{\sigma_{\rm s }^2 })}
{\rm d} \theta, 
\end{align*}
where  $F_U(u|\nu_0)=1-\int_{u}^{\infty}f_{U}(u|\nu_0) {\rm d} u$. Similarly, the PDF of $V_0$ can be obtained by substituting  \eqref{eq: Thomas location PDF} in \eqref{eq: PDF of V_0},  where $ V_0$ has a Rayleigh distribution with  scale parameter $\sigma_{\rm u}$.
\end{IEEEproof}
The conditional i.i.d. property of distances in the set   ${\cal D }^{{\bf x}_0}_{\rm s}$ enables us to characterize the distance from the typical user to its nearest open-access SBS located within the representative cluster. This result is presented in the next Lemma.
\begin{lemma} \label{Lem: density function of R_s} Conditioned on the  distance of the typical user to its cluster center, i.e. $V_0$, the CDF  of distance from  the typical user to its nearest open-access SBS, i.e., $R_{\rm s}$,  for a given $V_0=\nu_0$ is:
\begin{align} 
 F_{R_{\rm s}}(r_{\rm s}| \nu_0)=1-(1-F_{U}(r_{\rm s}|\nu_0))^{n_{{\rm s}_0}},\label{Eq: CDF closest small cell}
\end{align}
and the conditional PDF of  $R_{\rm s}$ for a given $\nu_0$ is:
\begin{align} \label{Eq: PDF closest small cell}
 f_{R_{\rm s}}(r_{\rm s}| \nu_0)=n_{{\rm s}_0}(1-F_{U}(r_{\rm s}|\nu_0))^{n_{{\rm s}_0}-1}  f_{U}(r_{\rm s}| \nu_0),
\end{align}
where  $F_{U}(\cdot|\nu_0)$ and $f_{U}(\cdot| \nu_0)$   are given by \eqref{eq: CDF of U} and \eqref{eq: PDF  of U}, respectively.
\end{lemma}
\begin{IEEEproof}
Conditioned on the distance of the typical user to its cluster center, the elements in ${\cal D }^{{\bf x}_0}_{\rm s}$ are i.i.d. with PDF $f_{U}(\cdot| \nu_0)$. Thus the result simply follows from the PDF of the minimum element of the i.i.d. sequence of random variables~\cite[eqn.~(3)]{david1970order}. 
\end{IEEEproof}
These distance distributions are  the keys to the derivation of the metrics of interest. 
\subsection{Association policies}
As discussed above, the candidate serving BS in each open-access tier (i.e., all macro BSs and SBSs located within the representative cluster) is the one nearest to the user. Recall that  the distances from a typical user to its  nearest open-access small cell and macro BSs were denoted by $R_{ \rm s}$ and ${R_{\rm m}}$, respectively. In order to select the serving BS from amongst the candidate serving BSs, we consider the following two association policies.
\subsubsection{Association Policy 1}
 The serving BS is chosen from amongst the candidate  serving BSs according to maximum  received-power averaged over small-scale fading. 
 As noted earlier,  this association policy maximizes coverage probability of a typical user.
 The association event to macro BSs and SBSs can be formally defined as follows.
\begin{itemize}
{\item A typical user is associated to a macrocell if $\argmax_{j\in\{{\rm s},{\rm m}\}} P_j R_j^{-\alpha}={\rm m}$. The association event to macrocell  is denoted by $S_{\rm m}^{\rm P1}$, where ${\bf 1}_{S_{\rm m}^{\rm P1}}={\bf 1}(\argmax_{j\in\{{\rm s},{\rm m}\}} P_j R_j^{-\alpha}={\rm m}).$}

\item A typical user is associated to a small cell if $\argmax_{j\in\{{\rm s},{\rm m}\}}  P_j R_j^{-\alpha}={\rm s}$. The association event to the small cell  is denoted by $S_{\rm s}^{\rm P1}$, where ${\bf 1}_{S_{\rm s}^{\rm P1}}={\bf 1}(\argmax_{j\in\{{\rm s},{\rm m}\}} P_j R_j^{-\alpha}={\rm s}).$
\end{itemize}
Now, the density functions of distances $f_{R_{\rm s}}(\cdot| \nu_0)$ and $f_{R_{\rm m}}(\cdot)$ obtained in  the previous subsection are used to characterize the association probabilities to macro and small cells in the next Lemma.

\begin{lemma} [Association probability for Policy 1] \label{lem : association prob}The  association probability of a typical user located at distance $\nu_0$ from its cluster center to the macrocell tier is:
\begin{align}
{\cal A}_{\rm m}^{\rm P1}(\nu_0)=\int_0^{\infty} \big[1-F_{R_{\rm s}}\big(\xi_{\rm s m } r_{\rm m}|\nu_0\big) \big] f_{R_{\rm m}}(r_{\rm m}) {\rm d} r_{\rm m},
\end{align}
 with $\xi_{\rm s m } =\big(\frac{P_{\rm s}}{P_{\rm m } }\big)^{1/\alpha}$, and the association probability to the  small cell tier is:
 \begin{align}
 {\cal A}_{\rm s}^{\rm P1}(\nu_0)=1-  {\cal A}_{\rm m}^{\rm P1}(\nu_0),
 \end{align}
 where $f_{R_{\rm m}}(\cdot)$ and  $F_{R_{\rm s}}(\cdot|\nu_0)$  are given by \eqref{Eq: PDF macro closest} and \eqref{Eq: CDF closest small cell}, respectively.
 \end{lemma}
\begin{IEEEproof}
See Appendix \ref{proof : association prob}.
\end{IEEEproof}
{The serving distance is simply the distance from the typical user to its nearest BS from associated tier. Denote by $X^{\rm P1}_j$ the serving distance to tier $j\in\{{\rm m},{\rm s}\}$. 
The density function of $X^{\rm P1}_j$ is characterized in the next Lemma.}

%
%
%
%
%
\begin{lemma} [Serving distance distribution under association  Policy 1] \label{lem: serving distance} For a  typical user located at distance $\nu_0$ from its cluster center, the PDF of serving distance $X_{\rm m}^{\rm P1}$ conditioned on the association to  macrocell, i.e., event $S_{\rm m}^{\rm P1}$, is:
\begin{align}
f_{X_{\rm m}^{\rm P1}} (x_{\rm m}|\nu_0)=\frac{1} {{\cal A}^{\rm P1}_{\rm m}(\nu_0)}(1-F_{R_{\rm s}}(\xi_{\rm s m  } x_{\rm m}|\nu_0)) f_{R_{\rm m}}(x_{\rm m}), \text{ and}
\end{align}
 the PDF of  $X_{\rm s}^{\rm P1}$ conditioned on the association to small cell, i.e., event $S_{\rm s}^{\rm P1}$, is:
\begin{align}
f_{X_{\rm s}^{\rm P1}} (x_{\rm s}|\nu_0)=\frac{1} {{\cal A}^{\rm P1}_{\rm s}(\nu_0)} (1-F_{R_{\rm m}}(\xi_{\rm  m s} x_{\rm s})) f_{R_{\rm s}}(x_{\rm s}|\nu_0),
\end{align}
where $\xi_{\rm s m } =\big(\frac{P_{\rm s} }{P_{\rm m } }\big)^{1/\alpha}$ and $\xi_{\rm m s}=\big(\frac{P_{\rm m} }{P_{\rm s} }\big)^{1/\alpha}$. The density functions  of distances $f_{R_{\rm m}}(\cdot)$,  $f_{R_{\rm s}}(\cdot|\nu_0)$, and $F_{R_{\rm s}}(\cdot|\nu_0)$ are given by  \eqref{Eq: PDF macro closest}, \eqref{Eq: PDF closest small cell}, and  \eqref{Eq: CDF closest small cell}, respectively.
\end{lemma}
\begin{IEEEproof}
See Appendix~\ref{Proof: serving distance}.
\end{IEEEproof}
From  association Policy 1,  it can be deduced that there {are} no open-access BSs within distance $\xi_{\rm m s} x_{\rm s}$ ($\xi_{\rm s m} x_{\rm m}$) of the typical user when this user   is served by small cell  (macrocell) BS. This can be interpreted as an exclusion zone with radius $\xi_{\rm m s} x_{\rm s}$ or $\xi_{\rm s m} x_{\rm m}$   {depending upon} the choice of serving BS, which is centered at the location of the typical user. The effect of exclusion zone on the interference caused by macro BSs (distributed according to a homogeneous PPP $\Phi_{\rm m}$) can be easily handled {using} the fact that  the distribution of the PPP conditioned on the location of  a point of PPP (here serving BS) is the same as that of the original PPP~\cite{haenggi2012stochastic}. However, characterizing the effect of  exclusion zone on the distribution of distances from clustered open-access interfering SBSs to the typical user  is more challenging. We define  the set ${\cal F}^{\rm P1}_{\rm s} ({\cal F }^{\rm P1}_{\rm m})$ to represent the sequence of distances from open access interfering SBSs (which by assumption belong to the representative cluster centered at ${\bf x}_0$) to the typical user conditioned on the serving BS belonging to small cell (macrocell), such that the elements of  $W^{\rm P1}_{\rm s} \in {\cal F }^{\rm P1}_{\rm s}$ and $W^{\rm P1}_{\rm m} \in {\cal F }^{\rm P1}_{\rm m}$ are greater than $\xi_{\rm m s} x_{\rm s}$  and $\xi_{\rm s m} x_{\rm m}$, respectively.
In the next Lemma, we deal with the conditional i.i.d. property of the elements of $ {\cal F }^{\rm P1}_{\rm s}$ and ${\cal F }^{\rm P1}_{\rm m}$, and their  distributions.
\begin{lemma}[Policy 1:  distribution of distances from open-access interfering SBSs] \label{Lem: Interfering distances from intra}  
Under association Policy 1, the elements  of ${\cal F }^{\rm P1}_{\rm m} \ ({\cal F }^{\rm P1}_{\rm s})$  conditioned on $V_0$ and $X^{\rm P1}_{\rm m} (X^{\rm P1}_{\rm s})$ are i.i.d., where the PDF of each element $W^{\rm P1}_{\rm m} \in {\cal F }^{\rm P1}_{\rm m} $ for a given  $V_0=\nu_0$ and $X^{\rm P1}_{\rm m}=x_{\rm m}$ is:
\begin{align}\label{eq: PDF Wm}
f_{W^{\rm P1}_{\rm m}}(w_{\rm m}|\nu_0,x_{\rm m})= \frac{f_{U}(w_{\rm m}|\nu_0)}{1-F_{U}( \xi_{\rm s m}x_{\rm m}|\nu_0)} ,
\end{align}
and the  conditional PDF of each element $W^{\rm P1}_{\rm s} \in {\cal F }^{\rm P1}_{\rm s}$ is:
\begin{align}\label{eq: PDF ws}
f_{W^{\rm P1}_{\rm s}}(w_{\rm s}|\nu_0,x_{\rm s})= \frac{f_{U}(w_{\rm s}|\nu_0)}{1-F_{U}(x_{\rm s}|\nu_0)},
\end{align}
where  $F_{U}(\cdot |\nu_0)$ and $f_{U}(\cdot|\nu_0)$ are given by \eqref{eq: CDF of U} and \eqref{eq: PDF  of U}, respectively.
\end{lemma}
\begin{IEEEproof}
See Appendix \ref{Proof: Interfering distances from intra}.
\end{IEEEproof}

\subsubsection{Association Policy 2}  
In addition to maximum RSRP-based association policy discussed above, it is often times desirable to define simple canonical association policies to  balance load across the network, which we do next. 
 For this purpose, the association event to the SBS and macro BS is defined as follows.


\begin{itemize}
\item A typical user is associated to the  small cell if  $ P_{\rm s} {R_{\rm s}}^{-\alpha} \ge P_0$.  The association event to the small cell is denoted by $S_{\rm s}^{\rm P2}$, where ${\bf 1}_{S_{\rm s}^{\rm P2}}={\bf 1}({R_{\rm s}}\le D )$ and $D\equiv \Big(\frac{P_{\rm 0}}{P_{\rm s}}\Big)^{-1/\alpha}$.

\item A typical user is associated to the macrocell if  $P_{\rm s} {R_{\rm s}}^{-\alpha} \le P_0$. The association event to the macrocell  is denoted by $S_{\rm m}^{\rm P2}$, where ${\bf 1}_{S_{\rm m}^{\rm P2}}={\bf 1}({R_{\rm s}}\ge D).$
\end{itemize}
Here $P_0$ denotes the SBS power threshold.
In contrast to the association Policy 1, which is a function of the distances from  both the nearest macro and small cell BSs to a typical user,  the  association Policy 2 is only a function of the distance of a typical user to its nearest open-access SBS, which lends relatively more tractability to the analysis. This simple policy allows us to balance load across macro and small cells by tuning the value of $P_0$. 
The exact impact of $P_0$ on the coverage probability will be studied in the later sections. According to the definition of $S_{\rm s}^{\rm P2}$, the conditional  association probability to the  small cell tier for a given $V_0=\nu_0$ is: 
\begin{align}\label{eq: association prob small cell distance}
{\cal A}_{\rm s}^{\rm P2}(\nu_0)=\E_{R_{\rm s}}[{\bf 1}(R_{\rm s} < D)|\nu_0]= \P(R_{\rm s} <D|\nu_0)= F_{R_{\rm s}}(D|\nu_0),
\end{align}
 and the association probability to the  macrocell tier is:
 \begin{align}
 {\cal A}^{\rm P2}_{\rm m}(\nu_0)=1-  {\cal A}_{\rm s}^{\rm P2}(\nu_0).
 \end{align}
Using the macro and small cell association probabilities, the density function of serving distance is derived in the next Lemma.
\begin{lemma}[Serving distance PDF for association Policy 2] \label{lem: serving distance distance based}
The PDF of serving distance $X^{\rm P2}_{\rm s}$ when the typical user located at distance $\nu_0$ from its own cluster center is served by a small cell is:
 \begin{align}
   f_{X^{\rm P2}_{\rm s}}(x_{\rm s}| \nu_0)= \frac{f_{R_{\rm s}}(x_{\rm s}| \nu_0)} {{\cal A}^{\rm P2}_{\rm s}(\nu_0)},  \quad 0\le x_{\rm s}\le D,
 \end{align}
 and the PDF of serving serving distance  $X^{\rm P2}_{\rm s}$ when the typical user is served by a macrocell is:
  \begin{align}
    f_{X^{\rm P2}_{\rm m}}(x_{\rm m})= {f_{R_{\rm m}}(x_{\rm m})},\quad x_{\rm m}\ge 0,
 \end{align}
where  ${f_{R_{\rm m}}(\cdot)}$, $f_{R_{\rm s}}(\cdot| \nu_0)$ and  ${{\cal A}^{\rm P2}_{\rm s}(\nu_0)}$ are given by \eqref{Eq: PDF macro closest}, \eqref{Eq: PDF closest small cell}, and \eqref{eq: association prob small cell distance}, respectively.
\end{lemma}
\begin{IEEEproof}  For the typical user located at distance $\nu_0$ from its own cluster center,    $f_{X^{\rm P2}_{\rm s}}(x_{\rm s}| \nu_0)$ is 
 the PDF of distance from the typical user to its nearest open-access SBS   conditioned on the association to small cell tier $S_{\rm s}^{\rm P2}$, which is equal to  $f_{R_{\rm s}}(x_{\rm s}|  S^{\rm P2}_{\rm s},\nu_0)$. However, the association to the macrocell is independent of the distance from the typical user to its nearest macro BS. Thus the PDF of serving distance when  the typical user is served by macrocell is simply the PDF of distance to its nearest macro BS.
\end{IEEEproof}
As has already been discussed earlier,  the locations of open-access interfering BSs depend upon the association event. From association Policy 2, it can be deduced that if a typical user is served by small cell (macrocell), the closest open-access interfering  SBS  must be at distance greater than $x_{\rm s}$ \Big($D\equiv \Big(\frac{P_{\rm 0}}{P_{\rm s}}\Big)^{-1/\alpha}$\Big) from the typical user. Denote by ${\cal F}^{\rm P2}_{\rm s}$ (${\cal F}^{\rm P2}_{\rm m}$) the sequence of distances from intra-cluster interfering SBSs to the typical user served by  small cell (macrocell).
The distributions of the elements of  $ {\cal F}^{\rm P2}_{\rm s}$  and $ {\cal F}^{\rm P2}_{\rm m}$ are given in the next Lemma.
\begin{lemma}[Association Policy 2: distribution of distances from open-access interfering SBSs]  \label{lem: Distance-based association intra-cluster interfering}
The elements in the sequence of distances from open-access interfering SBSs to the typical user served by macrocell, i.e., ${\cal F }^{\rm P2}_{\rm m}$, conditioned on $V_0$  are i.i.d., where the PDF of each element $W^{\rm P2}_{\rm m} \in {\cal F }^{\rm P2}_{\rm m}$  is:
\begin{align}\label{eq: PDF Wm distance based}
f_{W^{\rm P2}_{\rm m}}(w_{\rm m}|\nu_0)= \frac{f_{U}(w_{\rm m}|\nu_0)}{1-F_{U}(D|\nu_0)} ,
\end{align}
 and the elements  in the sequence of distances from open-access interfering SBSs to the typical user served by small cell, i.e., ${\cal F }^{\rm P2}_{\rm s}$,  are conditionally i.i.d.,  where the PDF of each element $W^{\rm P2}_{\rm s} \in {\cal F }^{\rm P2}_{\rm m}$ for  given $V_0=\nu_0$ and $X^{\rm P2}_{\rm s}=x_{\rm s}$ is:
\begin{align}\label{eq: PDF ws distance based}
f_{W^{\rm P2}_{\rm s}}(w_{\rm s}|\nu_0,x_{\rm s})= \frac{f_{U}(w_{\rm s}|\nu_0)}{1-F_{U}(x_{\rm s}|\nu_0)},
\end{align}
where $F_{U}(\cdot|\nu_0)$ and $f_{U}(\cdot |\nu_0)$ are given by \eqref{eq: CDF of U} and \eqref{eq: PDF  of U}, respectively.
\end{lemma}
\begin{IEEEproof}
The proof follows on the same lines as that of Lemma \ref{Lem: Interfering distances from intra}, and is hence skipped.
\end{IEEEproof}
The locations of  closed-access interfering SBSs are independent  of association policy. Thus the distribution of distances from  the typical user to the closed-access SBSs (also called inter-cluster interfering SBSs) is the same for  association policies~1 and 2. This distribution is presented next.
\begin{lemma} [Distribution of distances from closed-access interfering SBSs] \label{lem: PDF of inter-cluster interfering distance}
Denote by ${\cal D}^{\bf x}_{\rm s}=\{t_{\rm s}: t_{\rm s}=\|{\bf x}+{\bf y}_{\rm s}\|, \forall \ {\bf y}_{\rm s} \in {\cal N}^{\bf x}_{\rm s} \}$ the sequence of distances from the typical user to inter-cluster interfering SBSs within the cluster centered at ${\bf x}\in \Psi_{\rm p}$.
For a given $v= \|{\bf x}\|$, the elements of ${\cal D}^{\bf x}_{\rm s}$  are i.i.d., with PDF
\begin{align} \label{eq: PDF of inter-cluster interfering distance}
f_{T_{\rm s}} (t_{\rm s}| \nu)=\int_{-t_{\rm s}}^{t_{\rm s}} \frac{t_{\rm s}}{\sqrt{t_{\rm s}^2-z^2}} \Big[f_{{\bf Y}_{\rm s}}  \big(z-\nu, \sqrt{t_{\rm s}^2-z^2}\big)+ f_{{\bf Y}_{\rm s}}  \big(z-\nu, -\sqrt{t_{\rm s}^2-z^2}\big) \Big] {\rm d} z,\ t_{\rm s}>0.
\end{align}
\end{lemma}
\begin{IEEEproof}
The elements of the sequence ${\cal N}^{\bf x}_{\rm s}$, i.e., relative locations of the SBSs to the cluster centered at ${\bf x} \in {\Psi}_{\rm p}$, are i.i.d. by assumption. Hence, for a given $\nu=\|{\bf x}\|$, the elements of  the sequence  ${\cal D}^{\bf x}_{\rm s}$ are i.i.d. The derivation of $f_{T_{\rm s}} (\cdot | \nu)$ follows on the same lines  as that of $f_{U}(\cdot |\nu_0)$ given by \eqref{eq: PDF  of U}, and hence is skipped.
\end{IEEEproof}
\begin{remark}[Thomas cluster process]  For the special case of Thomas cluster process, the elements in the sequence of distances from closed-access interfering SBSs to the typical user are i.i.d., where the PDF of each element is: 
\begin{align}\label{eq: PDF of Ts, Rician}
 f_{T_{\rm s}}(t_{\rm s}|\nu)=\frac{t_{\rm s}}{ \sigma_{\rm s}^2} \exp\left(-\frac{t_{\rm s}^2+\nu^2}{2 \sigma_{\rm s}^2}\right) I_0\left(\frac{t_{\rm s} \nu}{\sigma_{\rm s}^2}\right), \quad t_{\rm s}>0,
\end{align}
which is Rician. The proof is exactly the same as that of Corollary~\ref{cor: PDFs  intra Thomas}. 
It should be noted that all the results can be specialized to   Thomas cluster process by substituting  $F_U(\cdot|\nu_0)$, $f_U(\cdot|\nu_0)$, $f_{V_0}(\cdot)$, and $f_{T_{\rm s}}(\cdot|\nu)$ with the expressions given by \eqref{eq: CDF of U, Rician}, \eqref{eq: PDF of U, Rician}, \eqref{eq: PDF of V_0, Rayleigh}, and 
\eqref{eq: PDF of Ts, Rician}, respectively.
\end{remark}
\section{Coverage Probability and  Throughput Analysis}
\label{sec: Coverage probability}
This is the second main technical section of this paper, where we use the distance distributions and association probabilities derived in the previous section to characterize network performance in terms of  coverage probability of a typical user and  throughput of the whole network. 
\subsection{Coverage probability}
The coverage probability  can be formally defined as the probability that   $\sir$  experienced by a typical user  is greater than  the desired threshold for successful demodulation and decoding. Mathematically, it is $\pc= \E[{\bf 1}(\sir>\beta)]=\P(\sir>\beta)$, where $\beta$ is the target $\sir$ threshold.
We specialize this definition to  the  two association policies   in this subsection. We begin our discussion with association Policy 1.
\subsubsection{Association Policy 1} 
As evident in the sequel,  the Laplace transform of (the PDF of)  interference  is the key intermediate result for the coverage probability analysis. Thus we first focus on the derivation of  the Laplace transform of  interference distribution.
As has already been described earlier, the contribution of the total interference seen at a typical user can be partitioned into three sources: i) interference caused by open-access SBSs, ii) interference caused by closed-access SBSs, and iii) interference caused by macro BSs. We now use the distance distributions  presented in Lemma~\ref{Lem: Interfering distances from intra} to characterize the Laplace transform of interference originating from the open-access SBSs (intra-cluster interferers) in the  next Lemma.
\begin{lemma} \label{lem: Laplace intra} Under  association Policy 1,
the  conditional  Laplace transform of interference  distribution caused by  open-access SBSs at a typical  user served by macrocell, ${{\cal I}^{\rm intra}_{\rm sm}}$,  for a  given  $X^{\rm P1}_{\rm m}= x_{\rm m}$ and $V_0=\nu_0$ is: 
\begin{align}\label{eq: Lap intra macro user}
{\cal L}^{\rm P1}_{{\cal I}^{\rm intra}_{\rm sm}}(s|\nu_0, x_{\rm m})=\sum_{\ell=0}^{n_{\rm s_0}} \left( \int_{ \xi_{\rm s m}x_{\rm m}}^{\infty}   \frac{1}{1+s P_{\rm s} w_{\rm m}^{-\alpha} } f_{W^{\rm P1}_{\rm m}}(w_{\rm m}|\nu_0,x_{\rm m})  \nrmd w_{\rm m} \right)^{\ell} \frac{\bar{n}_{\rm as}^\ell e^{-\bar{n}_{\rm as}}}{\ell! \sum_{k=0}^{n_{\rm s_0}} \frac{{\bar{n}_{\rm as}}^k e^{-\bar{n}_{\rm as}}}{k!}},
\end{align}
and the Laplace transform of interference distribution caused by  open-access SBSs at a typical user served by small cell, ${{\cal I}^{\rm intra}_{\rm ss}}$, is:
\begin{align}\label{eq: Lap intra small user}
{\cal L}^{\rm P1}_{{\cal I}^{\rm intra}_{\rm s s}}(s|\nu_0, x_{\rm s})=\sum_{\ell=1}^{n_{\rm s_0}} \left( \int_{x_{\rm s}}^{\infty}  \frac{1}{1+s P_{\rm s} w_{\rm s}^{-\alpha} }  f_{W^{\rm P1}_{\rm s}}(w_{\rm s}|\nu_0,x_{\rm s})  \nrmd w_{\rm s} \right)^{\ell-1} \frac{  {\bar{n}_{\rm as}}^{\ell-1} e^{-\bar{n}_{\rm as}}}{(\ell-1)! \sum_{k=1}^{n_{\rm s_0}} \frac{{\bar{n}_{\rm as}}^{k-1} e^{-\bar{n}_{\rm as}}}{(k-1)!}},
\end{align}
where $f_{W^{\rm P1}_{\rm m}}(\cdot|\nu_0, x_{\rm m})$ and $f_{W^{\rm P1}_{\rm s}}(\cdot|\nu_0, x_{\rm s})$ are given by Lemma~\ref{Lem: Interfering distances from intra}. 
\end{lemma}
\begin{IEEEproof}
See Appendix \ref{Proof: Laplace intra}.
\end{IEEEproof}
Recall that the total number of SBSs in the representative cluster was assumed to be  known {\em a priori} and equal to $n_{\rm  s_0}$. This assumption was made to simplify {\em order statistics} argument used in the derivation of the PDF of serving distance, but it  constrains  the maximum number of  interfering SBSs in the representative cluster, which complicates the numerical evaluation of the exact coverage probability (will be presented in Theorem~\ref{thm: coverage power based}) due to the  summation involved in the expressions given by Lemma~\ref{lem: Laplace intra}. However, these expressions can  be simplified under the assumption of $\bar{n}_{\rm as} \ll n_{{\rm s}_0}$, and the simplified expressions are presented in the next Corollary.
\begin{cor}\label{cor: Laplace intra power} For  $\bar{n}_{\rm as} \ll n_{{\rm s}_0}$,  the Laplace transform of interference given by \eqref{eq: Lap intra macro user} reduces to
\begin{align}
{\cal L}^{\rm P1}_{{\cal I}^{\rm intra}_{\rm sm}}(s|\nu_0, x_{\rm m})=\exp \Big( \bar{n}_{\rm as} \int_{ \xi_{\rm s m}x_{\rm m}}^{\infty}   \frac{1}{1+\frac {w_{\rm m}^{\alpha}}  {s P_{\rm s}}} f_{W^{\rm P1}_{\rm m}}(w_{\rm m}|\nu_0, x_{\rm m})  \nrmd w_{\rm m} \Big),
\end{align}
and the Laplace transform of interference given by \eqref{eq: Lap intra small user} reduces to
\begin{align}
{\cal L}^{\rm P1}_{{\cal I}^{\rm intra}_{\rm s s}}(s|\nu_0, x_{\rm s})=\exp\Big( \bar{n}_{\rm as} \int_{x_{\rm s}}^{\infty}  \frac{1}{1+ \frac{w_{\rm s}^{\alpha}}{s P_{\rm s}} }  f_{W^{\rm P1}_{\rm s}}(w_{\rm s}|\nu_0,x_{\rm s})  \nrmd w_{\rm s}\Big).
\end{align}
%
\end{cor}
For numerical evaluation, we will use the simpler expression presented in  Corollary~\ref{cor: Laplace intra power} instead of Lemma~\ref{lem: Laplace intra}. In the numerical results section, we will notice that the simplified expressions given  by  Corollary~\ref{cor: Laplace intra power} can be treated as a proxy of the exact expression for a wide range of cases.

Using the PDF of distance derived in Lemma \ref{lem: PDF of inter-cluster interfering distance}, we can derive the Laplace transform of interference distribution caused by closed-access interfering SBSs to the typical user (inter-cluster interference), which is stated in the next Lemma.
\begin{lemma}\label{lem: Laplace inter}
The Laplace transform of interference distribution from closed-access interfering SBSs to the typical user is:
\begin{align}\label{Eq: Lap_Inter}
{\cal L}_{{\cal I}^{\rm inter}_{{\rm s} j}}(s)= \exp\Big(-2 \pi \lambda_\nrmp \int_0^\infty \Big(1-\exp\Big(-  \bar{n}_{\rm as}  \int_0^\infty \frac{s P_{\rm s} {t_{\rm s}}^{-\alpha}}{1+s P_{\rm s} {t_{\rm s}}^{-\alpha}} f_{T_{\rm s}} (t_{\rm s}| \nu) {\rm d} t_{\rm s} \Big)\Big)\nu \nrmd \nu\Big),
\end{align}
where index $\ j\in \{m,s\}$ denotes the tier of the serving BS, and $f_{T_{\rm s}} (\cdot|\nu)$ is given by \eqref{eq: PDF of inter-cluster interfering distance}.
\end{lemma}
\begin{IEEEproof}
See Appendix \ref{Proof: Laplace inter}.
\end{IEEEproof}
After dealing with the interference from all open and closed-access SBSs,  we now focus on the Laplace transform of interference caused by macro BSs.
\begin{lemma} \label{lem: Laplace macro}The Laplace transform of interference from macro BSs (except the serving BS) at the typical user is:
\begin{align}\label{eq: Laplace macro}
{\cal L}_{{\cal I}_{{\rm m} j}}^{\rm P1} (s)=\exp\Big(-2 \pi \lambda_m  \int_{\xi_{{\rm m} j} x_j}^{\infty}  \frac{s P_{\rm m}u^{-\alpha}}{1+s P_{\rm m} u^{-\alpha}} u {\rm d} u \Big),
\end{align}
where index $j\in\{{\rm m}, {\rm s}\}$ denotes the choice of the serving BS.
\end{lemma}
\begin{IEEEproof}
The proof follows from that of \cite[Theorem 1]{jo2012heterogeneous} with a minor modification. 
\end{IEEEproof}
These Laplace transforms of interference distributions are used to evaluate  the coverage probability in the next Theorem.
\begin{theorem} [Coverage probability under association Policy 1] \label{thm: coverage power based}
Coverage probability of a typical user served by the $j^{th}$ tier is ${\tt P}^{\rm P1}_{{\rm c}_j} =\E[{\bf 1}\{\sir_j>\beta\}, {\bf 1}_{S_{j}^{\rm P1}} ]=$
\begin{align}\label{eq: per tier coverage power based}
&\int_0^{\infty} \int_0^{\infty} {\cal A}^{\rm P1}_j(\nu_0) {\cal L}^{\rm P1}_{{\cal I}^{\rm intra}_{{\rm s}j}} \Big( \beta \frac{  {x_j}^{\alpha}}{P_j } |\nu_0, x_j \Big)  {\cal L}_{{\cal I}^{\rm inter}_{{\rm s}j }}\Big( \beta \frac{  {x_j}^{\alpha}}{P_j } \Big) {\cal L}^{\rm P1}_{{\cal I}_{{\rm m }j}} \Big( \beta \frac{  {x_j}^{\alpha}}{P_j } \Big) f_{X^{\rm P1}_j} (x_j|\nu_0) f_{V_0} {(\nu_0)}  {\rm d} x_j {\rm d} \nu_0,
\end{align}
where $ j\in\{{\rm m},{\rm s}\}$. Using ${\tt P}^{\rm P1}_{{\rm c}_j}$, the total coverage probability is:
\begin{align}\label{eq: total coverage power based}
{\tt P}^{\rm P1}_{{\rm c}_{\rm T}}={\tt P}^{\rm P1}_{{\rm c}_{\rm m}}+{\tt P}^{\rm P1}_{{\rm c}_{\rm s}},
\end{align}
where $f_{V_0} {(\cdot)}$, ${\cal A}^{\rm P1}_j(\cdot)$, $ f_{X^{\rm P1}_j} (\cdot|\nu_0)$, ${\cal L}^{\rm P1}_{{\cal I}^{\rm intra}_{{\rm s}j}}(\cdot |\nu_0, x_{ j})$, ${\cal L}_{{\cal I}^{\rm inter}_{{\rm s}j}}(\cdot)$, and ${\cal L}^{\rm P1}_{{\cal I}_{{\rm m}j }}(\cdot)$ are given by Lemmas~\ref{lem: PDFs of intra-cluster distances}, \ref{lem : association prob}, \ref{lem: serving distance}, \ref{Lem: Interfering distances from intra}, \ref{lem: Laplace inter}, and \ref{lem: Laplace macro}, respectively. 
\end{theorem}
\begin{IEEEproof} The coverage probability of a typical user served by the $j^{th}$ tier is:
\begin{align*}
{\tt P}^{\rm P1}_{{\rm c}_j} &=           \E \Big[ {\bf1} \Big \{h_j \ge \beta \frac{  ( {X_j}^{\rm P1})^{\alpha}}{P_j } ({\cal I}_{{\rm m}j}+{\cal I}_{{\rm s}j}^{\rm intra}+{\cal I}_{{\rm s}j}^{\rm inter}) \Big\} ,  {\bf 1}_{S_{j}^{\rm P1}}  \Big]   \\
&\stackrel{}{=}\E \Big[\P\Big(h_j \ge \beta \frac{ ( {X_j}^{\rm P1})^{\alpha}}{P_j } ({\cal I}_{{\rm m}j}+{\cal I}_{{\rm s}j}^{\rm intra}+{\cal I}_{{\rm s}j}^{\rm inter}) \Big | S_j^{\rm P1}, V_0\Big) \P \Big(S_j^{\rm P1}|V_0\Big)\Big]\\
&\stackrel{(a)}{=}\E \Big[\E \Big[\exp\Big( \beta \frac{  ( {X_j}^{\rm P1})^{\alpha}}{P_j } ({\cal I}_{{\rm m}j}+{\cal I}_{{\rm s}j}^{\rm intra}+{\cal I}_{{\rm s}j}^{\rm inter}) \Big)\Big | S_j^{\rm P1}, V_0\Big] {\cal A}^{\rm P1}_j(V_0)\Big],
\end{align*}
%
where  $(a)$ follows from Rayleigh fading assumption, i.e., $h_j\sim \exp(1)$ and association probability definition. The final expression of  ${\tt P}^{\rm P1}_{{\rm c}_j} $ given by \eqref{eq: per tier coverage power based} is obtained by using the definition of Laplace transform along with independence of open-access (intra-cluster), closed-access (inter-cluster) SBSs, and macro BSs interference powers, followed by de-conditioning over  $X^{\rm P1}_j$ given $V_0=\nu_0$, followed by de-conditioning over $V_0$. Now using ${\tt P}^{\rm P1}_{{\rm c}_j}$, the total coverage probability is obtained by applying the law of total probability.
\end{IEEEproof}
\subsubsection{Association Policy 2}
We  extend the coverage probability analysis  to the case where the serving BS is chosen according to  association Policy 2.  Similar to the previous subsection, we begin by deriving the Laplace transform of interference distribution. Using the PDF of distance derived in Lemma  \ref{lem: Distance-based association intra-cluster interfering}, the Laplace transform of interference caused by open-access  interfering SBSs is characterized in the next Lemma.
\begin{lemma}\label{lem: Laplace intra distance based}
Under association Policy 2, the Laplace transform of interference  caused by  open-access SBSs at a typical user served by macrocell, ${{\cal I}^{\rm intra}_{\rm sm}}$, conditioned on $V_0$ is: 
\begin{align}
{\cal L}^{\rm P2}_{{\cal I}^{\rm intra}_{\rm sm}}(s|\nu_0)=  \sum_{\ell=0}^{n_{\rm s_0}} \left( \int_{ D}^{\infty}   \frac{1}{1+s P_{\rm s} w_{\rm m}^{-\alpha} } f_{W^{\rm P2}_{\rm m}}(w_{\rm m}|\nu_0)  \nrmd w_{\rm m} \right)^{\ell}  \frac{\bar{n}_{\rm as}^\ell e^{-\bar{n}_{\rm as}}}{\ell! \sum_{k=0}^{n_{\rm s_0}} \frac{{\bar{n}_{\rm as}}^k e^{-\bar{n}_{\rm as}}}{k!}},
\end{align}
which for $\bar{n}_{\rm as} \ll n_{{\rm s}_0}$ simplifies to
\begin{align}
{\cal L}^{\rm P2}_{{\cal I}^{\rm intra}_{\rm sm}}(s|\nu_0)=\exp \Big( \bar{n}_{\rm as} \int_{ D}^{\infty}   \frac{1}{1+\frac {w_{\rm m}^{\alpha}}  {s P_{\rm s}}} f_{W^{\rm P2}_{\rm m}}(w_{\rm m}|\nu_0)  \nrmd w_{\rm m} \Big),
\end{align}
where $f_{W^{\rm P2}_{\rm m}}(\cdot|\nu_0)$ is given by \eqref{eq: PDF Wm distance based}.
The Laplace transform of interference  caused by  open-access SBSs at a typical small cell user is the same for  the two association policies. Thus we have ${\cal L}^{\rm P2}_{{\cal I}^{\rm intra}_{\rm s s}}(s|\nu_0, x_{\rm s})={\cal L}^{\rm P1}_{{\cal I}^{\rm intra}_{\rm s s}}(s|\nu_0, x_{\rm s})$, where ${\cal L}^{\rm P1}_{{\cal I}^{\rm intra}_{\rm s s}}(\cdot)$ is given by~\eqref{eq: Lap intra small user}.
\end{lemma}
\begin{IEEEproof}
The proof follows on the same lines  as that of Lemma \ref{lem: Laplace intra}, where the nearest open-access SBS is located at distance greater than $D$  to the typical user served by macrocell.
\end{IEEEproof}

As noted above, the interference caused by closed-access SBSs is independent of association policy, and hence its Laplace transform  is the same for the two association policies.
Now we are left with the derivation  of the Laplace transform of interference caused by macro BSs, which is presented in the next Lemma. 
\begin{lemma}\label{lem : laplace macro distance based}
The Laplace transform of interference from macro BSs  at a typical  user served by small cell is:
\begin{align}
{\cal L}^{\rm P2}_{{\cal I}_{{\rm ms} }} (s)= \exp\Bigg(-\pi \lambda_{\rm m}\frac{(s P_{\rm m})^{\frac{2}{\alpha}}}{\sinc(\frac{2}{\alpha})}\Bigg),
\end{align}
and the Laplace transform of interference  from macro BSs (except serving) at a typical user served by macrocell is the same for the two association policies. 
Thus we have ${\cal L}^{\rm P2}_{{\cal I}_{{\rm m m} }} (s)={\cal L}^{\rm P1}_{{\cal I}_{{\rm m m} }} (s)$, where ${\cal L}^{\rm P1}_{{\cal I}_{{\rm m m} }} (\cdot)$ is given by~\eqref{eq: Laplace macro}.
\end{lemma}
\begin{IEEEproof}
The proof follows on the same lines  as that of Lemma \ref{lem: Laplace macro}. The main difference is that  association Policy 2 imposes no constraint on the location of interfering macro BSs to the typical user served by the small cell. Thus we have
\begin{align*}
{\cal L}^{\rm P2}_{{\cal I}_{{\rm ms} }} (s)=\exp\Big(-2 \pi \lambda_m  \int_{0}^{\infty}  \frac{s P_{\rm m} u^{-\alpha}}{1+s P_{\rm m} u^{-\alpha}} u {\rm d} u \Big),
\end{align*}
where the
final expression is obtained by using \cite[(3.241)]{zwillinger2014table}.
\end{IEEEproof}
Using these Lemmas, we now derive the  coverage probability of a typical user for association Policy 2. The proof follows on the same lines  as that of Theorem~\ref{thm: coverage power based}.
\begin{theorem} [Coverage probability under association Policy 2]  \label{thm: coverage distance}
The coverage probability of a typical user served by small cell is: ${\tt P}^{\rm P2}_{{\rm c}_{\rm s}}=\E[{\bf 1}\{\sir_{\rm s}>\beta\}, {\bf 1}_{S_{\rm s}^{\rm P2}}]=$
\begin{align}
 \int_0^{\infty} \int_0^{D}  {\cal A}^{\rm P2}_{\rm s}(\nu_0) {\cal L}^{\rm P2}_{{\cal I}^{\rm intra}_{{\rm s s}}}( \beta \frac{  {x_{\rm s}}^{\alpha}}{P_{\rm s} } |\nu_0, x_{\rm s})  {\cal L}_{{\cal I}^{\rm inter}_{{\rm s s} }}( \beta \frac{  {x_{\rm s}}^{\alpha}}{P_{\rm s} } ) {\cal L}^{\rm P2}_{{\cal I}_{{\rm m s }}} ( \beta \frac{ {x_{\rm s}}^{\alpha}}{P_{\rm s} } ) f_{X^{\rm P2}_{\rm s}} (x_{\rm s}|\nu_0) f_{V_0} {(\nu_0)} 
{\rm d} x_{\rm s} {\rm d} \nu_0,
\end{align}
and the coverage probability of a typical user served by macrocell is: 
\begin{align} \notag
{\tt P}^{\rm P2}_{{\rm c}_{\rm m}} = \E[{\bf 1}\{\sir_{\rm m}>\beta\}, {\bf 1}_{S_{\rm m}^{\rm P2}}]&=
\int_0^{\infty}\int_0^{\infty}   {\cal A}^{\rm P2}_{\rm m}(\nu_0) {\cal L}^{\rm P2}_{{\cal I}^{\rm intra}_{\rm sm}}( \beta \frac{  {x_{\rm m}}^{\alpha}}{P_{\rm m} }  |\nu_0)  {\cal L}_{{\cal I}^{\rm inter}_{{\rm s m} }}( \beta \frac{  {x_{\rm m}}^{\alpha}}{P_{\rm m} } ) \\
&\times    {\cal L}^{\rm P2}_{{\cal I}_{{\rm m m} }} (\beta \frac{  {x_{\rm m}}^{\alpha}}{P_{\rm m} } ) f_{X^{\rm P2}_{\rm m}} (x_{\rm m}) f_{V_0} {(\nu_0)} {\rm d} x_{\rm m}  {\rm d} \nu_0,
\end{align}
using which the total coverage probability is:
\begin{align}\label{eq:  coverage power total distance}
{\tt P}^{\rm P2}_{{\rm c}_{\rm T}} ={\tt P}^{\rm P2}_{{\rm c}_{\rm m}} + {\tt P}^{\rm P2}_{{\rm c}_{\rm s}},
\end{align}
where $f_{V_0} {(\cdot)}$, $f_{X^{\rm P2}_j} (\cdot)$, ${\cal L}^{\rm P2}_{{\cal I}^{\rm intra}_{{\rm s j}}}(\cdot)$, ${\cal L}_{{\cal I}^{\rm inter}_{{\rm s j} }}(\cdot)$, and ${\cal L}^{\rm P2}_{{\cal I}_{{\rm m j} }} (\cdot)$ are given by  Lemmas~\ref{lem: PDFs of intra-cluster distances},    \ref{lem: serving distance distance based}, \ref{lem: Laplace intra distance based}, \ref{lem: Laplace inter}, and  \ref{lem : laplace macro distance based}, respectively. 
\end{theorem}
\begin{remark}[Optimal SBS power threshold $P_0$]
Decreasing SBS power threshold  has  a conflicting effect on the association to macrocell and small cell:  association probability to macrocell decreases whereas association probability to small cell increases.  In the Numerical Results Section, we concretely demonstrate  that there exists an optimal SBS power threshold $P_0$ (or equivalently distance threshold $D$) that maximizes the total coverage probability. Similarly, optimal $P_0$ can also be determined to balance load across macro and small cells so as to maximize the overall rate coverage probability. Further investigation on the rate coverage and load balancing is left as a promising future direction.
\end{remark}
Using these coverage probability results, we characterize  throughput  in the next subsection.
\subsection{Throughput}
In order to study the tradeoff  between aggressive frequency reuse and resulting interference, we use the following notion of the network {\em throughput}~\cite{ThroughputQuek2012}: 
\begin{align}
{\cal T}= \lambda \log_2(1+\beta) \pc,
\end{align}
where $\lambda$ is the number of simultaneously active transmitters per unit area. This metric roughly characterizes the average
number of bits successfully transmitted per unit area.
This definition is specialized to our setup in the next Proposition.
\begin{prop}Using the result of Theorem \ref{thm: coverage power based}, the throughput of association Policy~1 is:
\begin{align}
{\cal T}^{\rm P1}= (\lambda_{\rm m}{\tt P}^{\rm P1}_{{\rm c}_{\rm m}}  +  \lambda_{\rm p}  \bar{n}_{\rm as}  {\tt P}^{\rm P1}_{{\rm c}_{\rm s}})\log_2(1+\T), 
\end{align}
and using the result of Theorem \ref{thm: coverage distance}, the throughput of association Policy~2 is:
\begin{align}
{\cal T}^{\rm P2}=( \lambda_{\rm m}{\tt P}^{\rm P2}_{{\rm c}_{\rm m}} +  \lambda_{\rm p}  \bar{n}_{\rm as} {\tt P}^{\rm P2}_{{\rm c}_{\rm s}}) \log_2(1+\T).  
\end{align}
\end{prop}
\begin{remark}[Number of simultaneously active SBSs within  a cluster] \label{remark: Number of active SBS}
Increasing the number of simultaneously active SBSs  boosts the spectral efficiency by more aggressive frequency reuse,  whilst it leads to higher interference power.
While it is straightforward to conclude from the analytical results that the coverage probability always decreases with the number of simultaneously active SBSs, we will demonstrate in the next section that the throughput increases with the number of simultaneously active SBSs in the regime of interest. This in turn implies that the usual assumption of strictly orthogonal channelization per cluster, {i.e.,  only one simultaneously active SBSs per cluster} (e.g., see \cite{HetPCPGhrayeb2015}), should be revisited.
\end{remark}
\section{Results and Discussion} \label{sec:NumResults}
\subsection{Verification of results}
\begin{figure}
 \begin{minipage}{.49\textwidth}
  \includegraphics[width=1\textwidth]{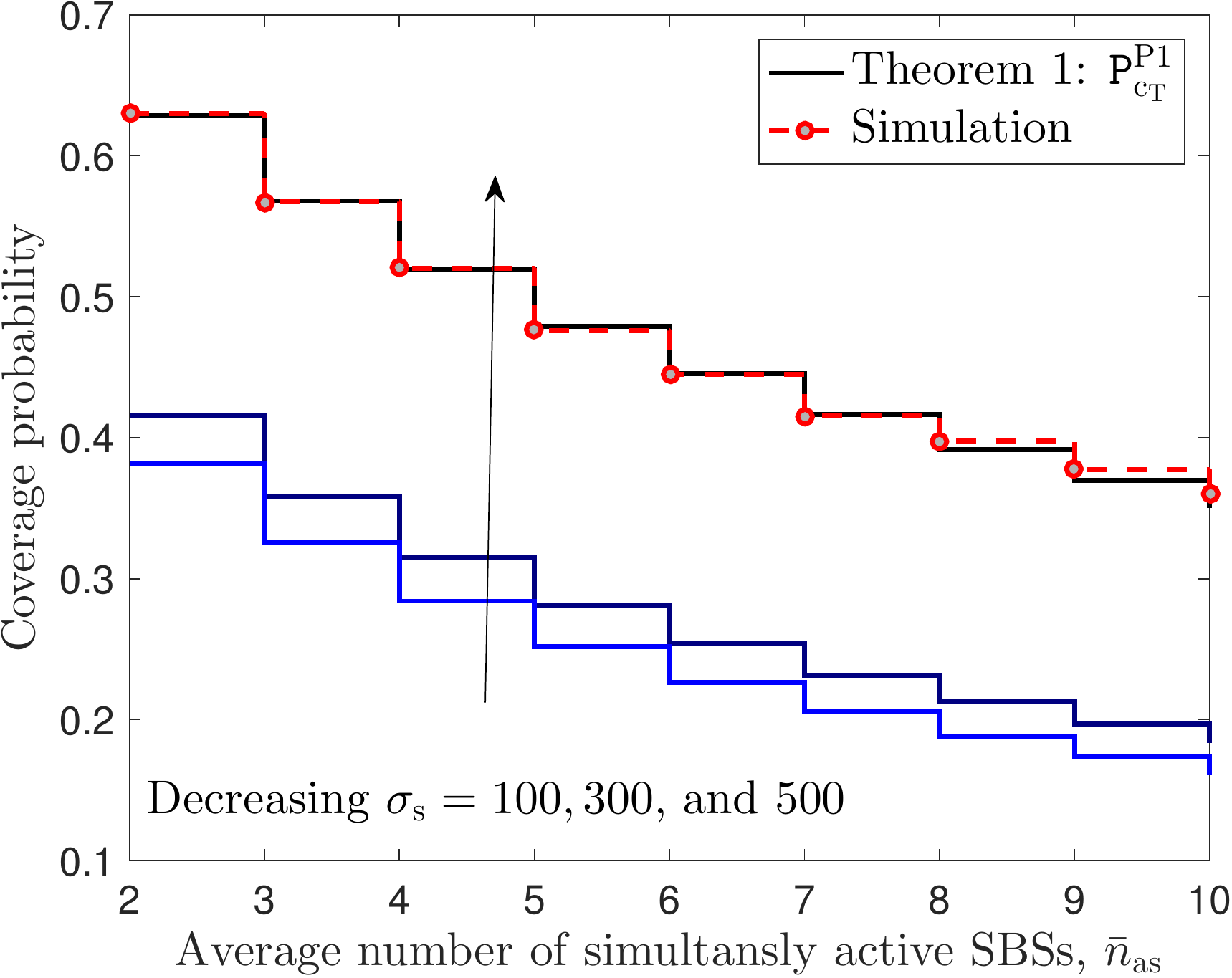}
              \caption{ Coverage probability as a function of average number of simultaneously active SBSs for different value
of $\sigma_{\rm s}$ under  association Policy~1
               ($n_{{\rm s}_0}=10$ and $\sigma_{\rm s}=\sigma_{\rm u}$)}
                \label{Fig: Cov_P_based_number_active}
\end{minipage}%
\hfill
 \begin{minipage}{.49\textwidth}

  \includegraphics[width=1\textwidth]{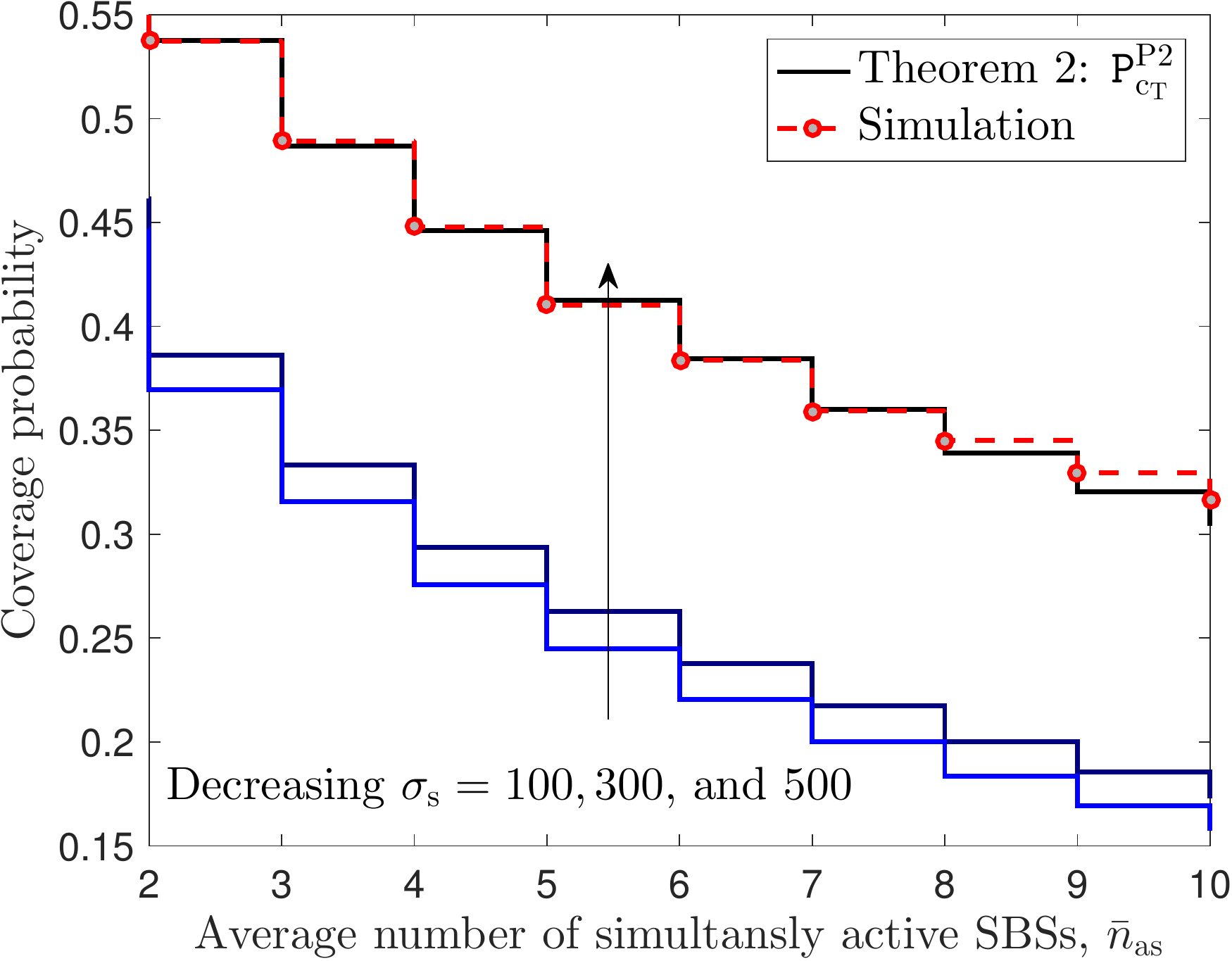}
              \caption{ Coverage probability as a function of average number of simultaneously active SBSs for different value
of $\sigma_{\rm s}$ under association Policy~2
               ($n_{{\rm s}_0}=10$ and $\sigma_{\rm s}=\sigma_{\rm u}$)}
                \label{Fig: Cov_D_based_number_active}
\end{minipage}
\end{figure}

In this section, we verify the accuracy of the  analysis by  comparing the analytical results with Monte Carlo simulations.
For this comparison,  the macro BS locations  are distributed as an independent PPP with density $\lambda_{\rm m}=1$ km$^{-2}$, and the geographical centers of user hotspots (i.e., cluster centers) are distributed as an independent PPP with density $\lambda_{\rm p}=10$ km$^{-2}$ around which users and SBSs are assumed to be  normally distributed with variances $\sigma_{\rm u}^2$ and $\sigma_{\rm s}^2$, respectively. 
 For this setup, we set the path-loss exponent, $\alpha$ as 4, the $\sir$ threshold as 0 dB, power ratio $P_{\rm m}=10^3 P_{\rm s}$,  $P_{\rm s}=23$ dBm, and  study the coverage probability for the two association policies.  As discussed in Section \ref{sec: Coverage probability},  the summation involved in the exact expression of Laplace transform of  intra-cluster interference complicates the numerical evaluation of Theorems \ref{thm: coverage power based} and \ref{thm: coverage distance}. Thus we use simpler expressions of Laplace transform of intra-cluster interference derived under the assumption $\bar{n}_{\rm as} \ll n_{{\rm s}_0}$ presented in Corollary~\ref{cor: Laplace intra power} and Lemma~\ref{lem: Laplace intra distance based}  for numerical evaluation of Theorems \ref{thm: coverage power based} and \ref{thm: coverage distance}. As evident from 
 Figs. \ref{Fig: Cov_P_based_number_active} and \ref{Fig: Cov_D_based_number_active}, the simpler expressions  can be treated as proxies for the exact  ones for wide range of parameters.  Considering $n_{{\rm s}_0}=10$,  the analytical plots exhibit  perfect match with simulation even for relatively large values of $\bar{n}_{\rm as}$.
Comparing Figs~\ref{Fig: Cov_P_based_number_active} and \ref{Fig: Cov_D_based_number_active}, we  also note that 
 the coverage probability   for association Policy~1 is higher than that of  Policy~2.

 \begin{figure}
 \begin{minipage}{.49\textwidth}
 
   \includegraphics[width=1\textwidth]{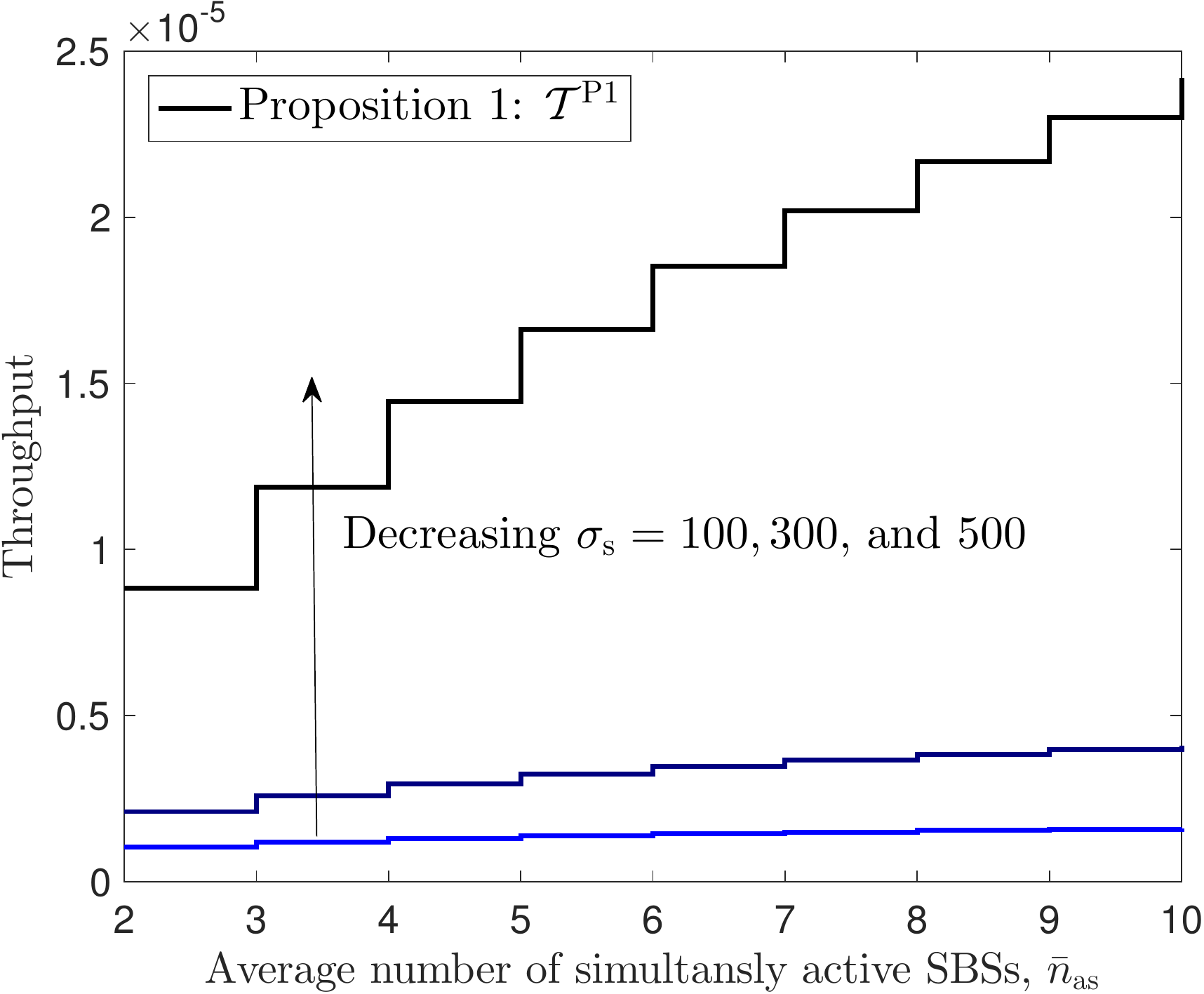}
              \caption{ Throughput as a function of average number of simultaneously active SBSs for different value
of $\sigma_{\rm s}$ under  association Policy~1
               ($n_{{\rm s}_0}=10$ and $\sigma_{\rm s}=\sigma_{\rm u}$)}
\label{Fig:  ASE_P_based_number_active }
\end{minipage}%
\hfill
 \begin{minipage}{.49\textwidth}
  \includegraphics[width=1\textwidth]{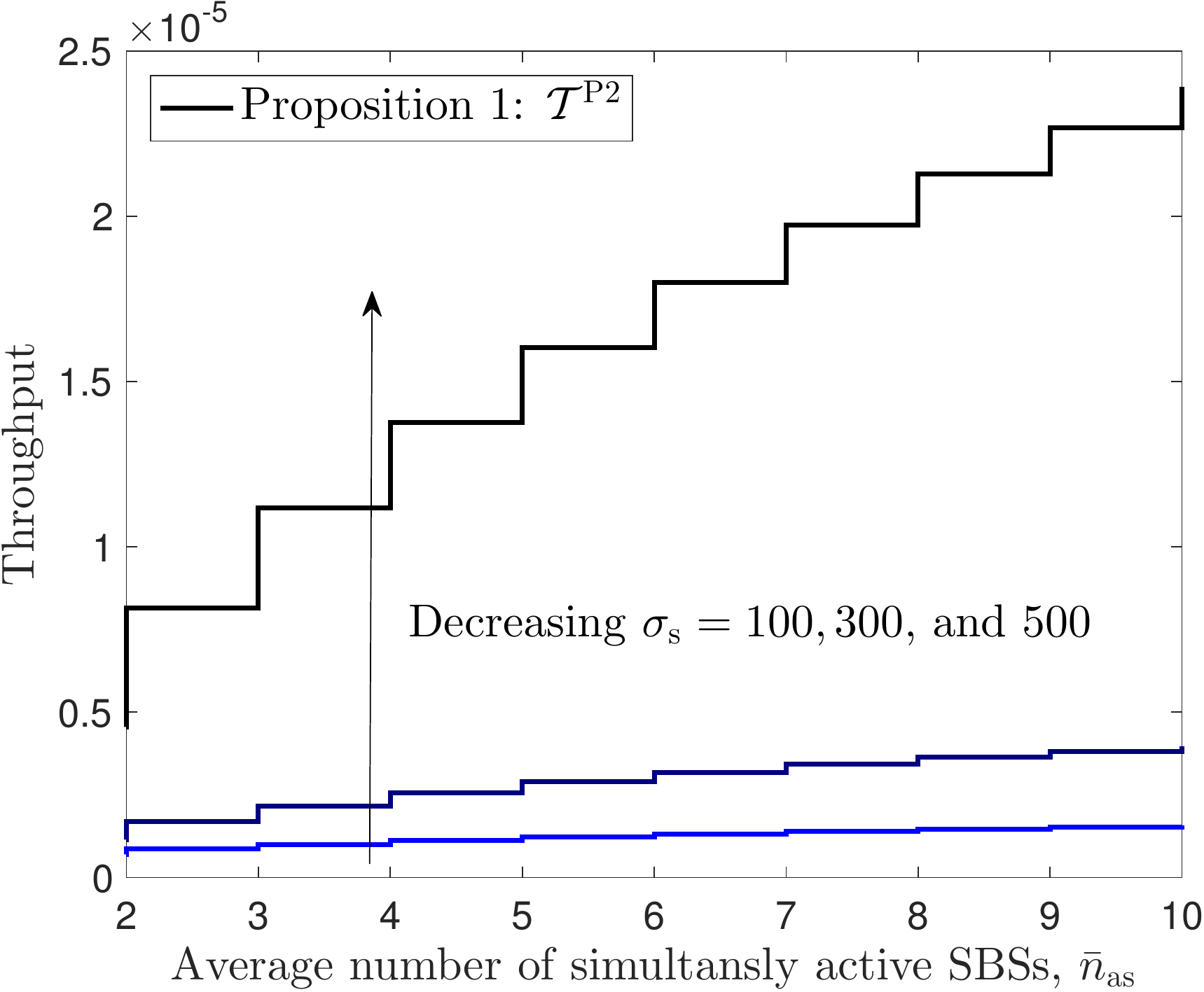}
              \caption{ Throughput as a function of average number of simultaneously active SBSs for different value
of $\sigma_{\rm s}$ under  association Policy~2
               ($n_{{\rm s}_0}=10$ and $\sigma_{\rm s}=\sigma_{\rm u}$)}
\label{Fig:  ASE_D_based_number_active}
\end{minipage}
\end{figure}

\subsection{Number of simultaneously active SBSs}
The coverage probabilities as  a function  of average number of simultaneously active SBSs, $\bar{n}_{\rm as}$, are presented in  Figs. \ref{Fig: Cov_P_based_number_active} and \ref{Fig: Cov_D_based_number_active} for association policies 1 and 2.  Our analysis
concretely demonstrates that the coverage probability always  decreases when more SBSs per cluster reuse the same spectrum.
 This is because having more simultaneously active SBSs results in more interference. However, there is a classical trade-off between frequency reuse and resulting interference. To study this trade-off, we plot  throughput as a function of $\bar{n}_{\rm as}$ in Figs. \ref{Fig:  ASE_P_based_number_active } and \ref{Fig:  ASE_D_based_number_active}. Interestingly in the considered range,  throughput increases with  the average number of simultaneously active SBSs per cluster. This means that more and more SBSs can be simultaneously activated as long as the coverage probability remains acceptable. From this observation, it can also be deduced that strictly orthogonal channelization (at most one SBS is allowed to use a given time-frequency resource element/block per cluster) is not spectrally efficient.

 \subsection{Impact of  SBS standard deviation}
\begin{figure}
\begin{minipage}{.49\textwidth}
  \includegraphics[width=1\textwidth]{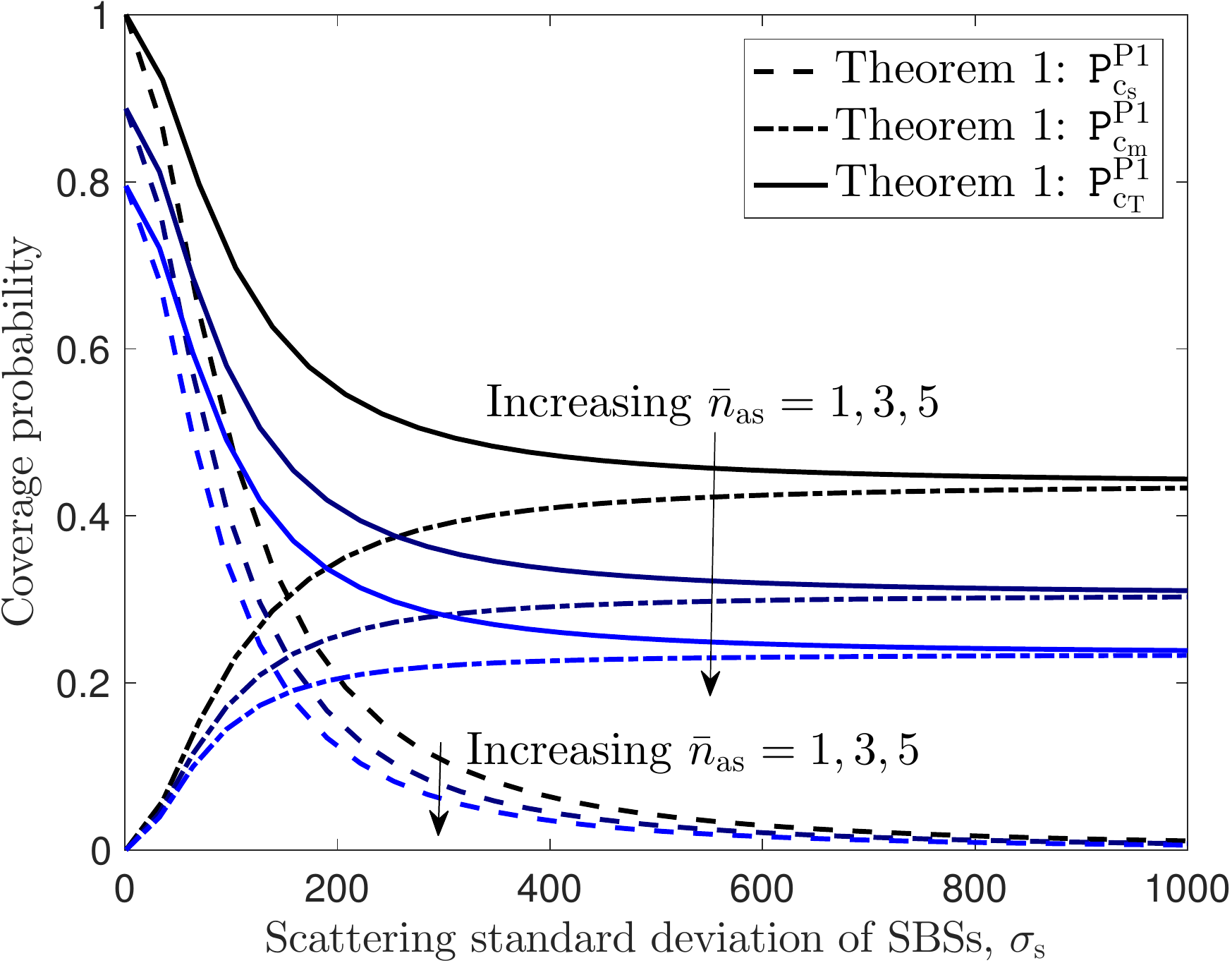}
              \caption{Coverage probability as a function of standard deviation $\sigma_{\rm s}$ for different value of $\bar{n}_{\rm a s}$
               ($n_{{\rm s}_0}=10$ and $\sigma_{\rm s}=\sigma_{\rm u}$)}
\label{Fig:  Cov_P_based_SigmaS}
\end{minipage}
\hfill
  \begin{minipage}{.49\textwidth}
  \includegraphics[width=1\textwidth]{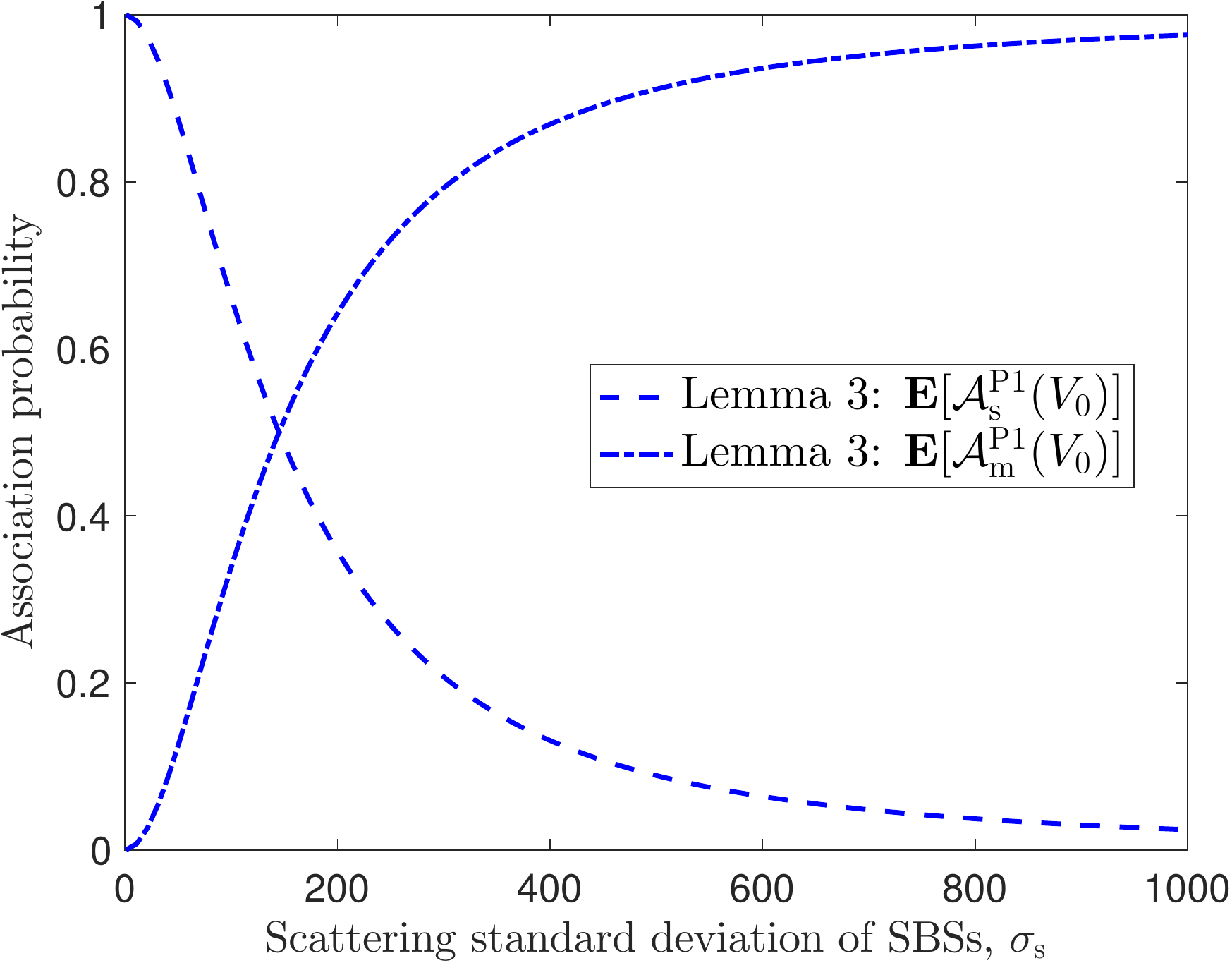}
              \caption{ Association probability as a function of standard deviation $\sigma_{\rm s}$
               ($n_{{\rm s}_0}=10$ and $\sigma_{\rm s}=\sigma_{\rm u}$)}
                \label{Fig: Association_P_based_Sigma}
\end{minipage}%

\end{figure}

For association Policy~1, the coverage probability  as  a function  of scattering standard deviation of SBSs, $\sigma_{\rm s}$, is plotted in \figref{Fig:  Cov_P_based_SigmaS}. The plot shows that  $\sigma_{\rm s}$ has a conflicting effect on ${\tt P}_{{\rm c}_{\rm m}}^{\rm P1}$ and ${\tt P}_{{\rm c}_{\rm s}}^{\rm P1}$: ${\tt P}_{{\rm c}_{\rm m}}^{\rm P1}$ increases and  ${\tt P}_{{\rm c}_{\rm s}}^{\rm P1}$ decreases.  The intuition behind this observation is that by increasing  $\sigma_{\rm s}$ the association to macrocell increases while association to small cell decreases. In  \figref{Fig: Association_P_based_Sigma}, we  plot average association probability as a function of  $\sigma_{\rm s}$ to exhibit this trend. From \figref{Fig:  Cov_d_base_sigma}, a similar observation can be made for  association Policy~2, where ${\tt P}_{{\rm c}_{\rm m}}^{\rm P2}$ increases and  ${\tt P}_{{\rm c}_{\rm s}}^{\rm P2}$ decreases as scattering standard deviation of SBSs increases.

%
%
%
%
%

%

\subsection{Optimal distance threshold}
 
\begin{figure}
 \begin{minipage}{.49\textwidth}
  \includegraphics[width=1\textwidth]{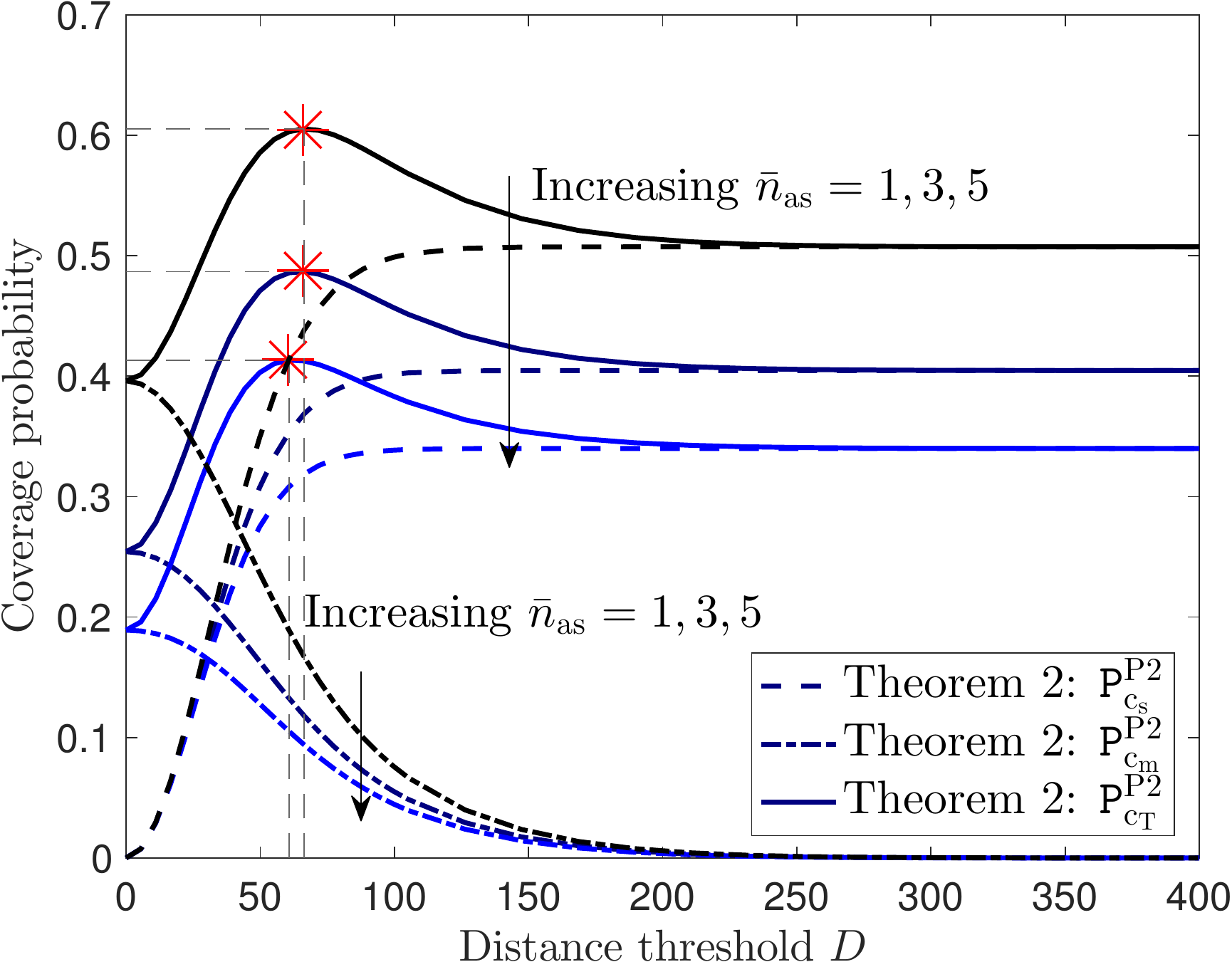}
              \caption{Coverage probability  as  a function of ~$D\equiv \Big(\frac{P_{\rm 0}}{P_{\rm s}}\Big)^{-1/\alpha}$ for different value of~$\bar{n}_{\rm a s}$
               ($n_{{\rm s}_0}=10$ and $\sigma_{\rm s}=\sigma_{\rm u}$)}
                \label{Fig: Cov_d_base_active_SBS}
\end{minipage}%
\hfill
 \begin{minipage}{.49\textwidth}
  \includegraphics[width=1\textwidth]{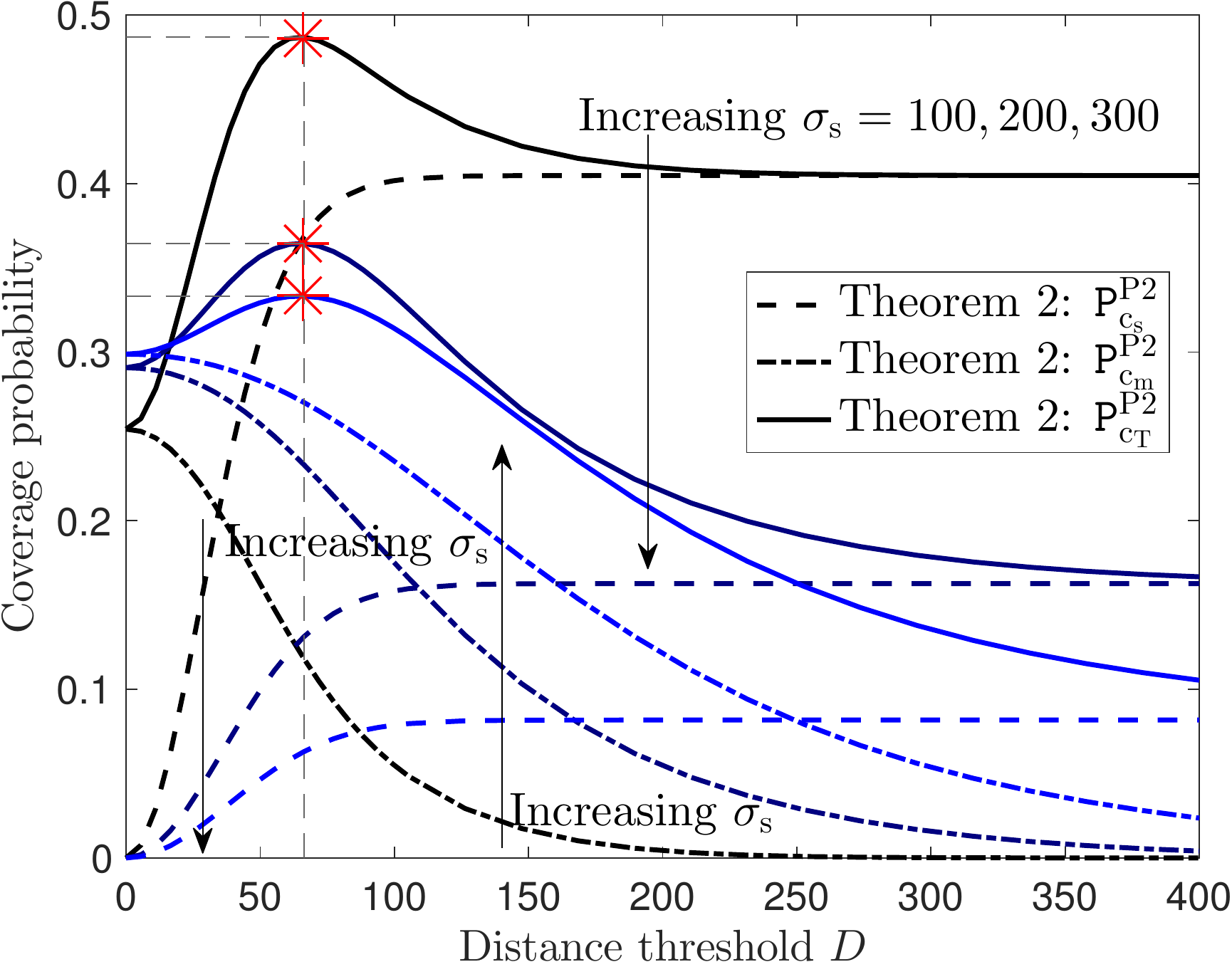}
              \caption{Coverage probability as a function of  $D\equiv \Big(\frac{P_{\rm 0}}{P_{\rm s}}\Big)^{-1/\alpha}$ for different value of $\sigma_{\rm s}$
               ($n_{{\rm s}_0}=10$  and $\sigma_{\rm s}=\sigma_{\rm u}$).}
\label{Fig:  Cov_d_base_sigma}
\end{minipage}
\end{figure}

 As evident from Figs. \ref{Fig: Cov_d_base_active_SBS} and \ref{Fig:  Cov_d_base_sigma}, there exists  an optimal SBS power  threshold \big(or equivalently distance threshold  $D\equiv \big(\frac{P_{\rm 0}}{P_{\rm s}}\big)^{-1/\alpha}$\big) that maximizes the total coverage probability.  The existence of the optimal value can be intuitively justified by the conflicting effect of the power threshold on the association to macrocell and small cell, as discussed in Remark~\ref{remark: Number of active SBS}. 
   From  \figref{Fig: Cov_d_base_active_SBS}, we can observe that the optimal distance threshold decreases with the increase of  average number of simultaneously active SBSs per cluster. This is because although both ${\tt P}_{{\rm c}_{\rm s}}^{\rm P2}$ and ${\tt P}_{{\rm c}_{\rm m}}^{\rm P2}$ decrease with the increase of $\bar{n}_{\rm a s}$, the former decreases at a slightly higher rate.
 Thus it is desirable to associate less users to SBSs.  Interestingly, we notice that the optimal distance threshold for different values of $\sigma_{\rm s}$ does not change in the setup considered in \figref{Fig:  Cov_d_base_sigma}.

%
%

%

\section{Concluding Remarks}
We developed a comprehensive framework for the performance analysis of HetNets with user-centric capacity-driven small cell deployments. Unlike the prior art on the spatial modeling of HetNets where users and SBSs are usually modeled by independent homogeneous PPPs, we introduced a tractable approach to incorporate coupling in the locations of the users and SBSs in HetNets, which bridges the gap between the simulation models used by industry (especially for the user hotspots), such as by 3GPP~\cite{access2010further}, and the ones used thus far by the stochastic geometry community. In particular, we assumed that  the geographical centers of user hotspots are distributed according to a homogeneous PPP around which users and SBSs are located with two general distributions. This approach not only models the aforementioned coupling, but also captures the non-homogeneous nature of user distributions~\cite{access2010further}. For this setup, we derived the coverage probability of a typical user and throughput of the whole network for two received power-based association policies. Our setup is general and applicable to any distributions of the relative locations of the users and SBSs with respect to the cluster center. A key intermediate step is the derivation of a new set of distance distributions, which
enabled the accurate analysis of user-centric small cell deployments. For numerical evaluation, we considered the special case of Thomas cluster process, which led to several design insights. The most important one is that the throughput increases with the increase of the number of SBSs per cluster reusing the same resource block in the considered setup. Therefore, the usual assumption of strictly orthogonal channelization per cluster should be revisited for efficient design, planning, and dimensioning of the system.
The proposed approach has numerous extensions. From modeling perspective, it is desirable to develop a {\em unified} analytical model to encompass different  spatial configurations considered by 3GPP for modeling BS and user locations in HetNets~\cite{SahaAfshDhiITA2017}. Further, it is desirable to choose point process models that simultaneously capture the spatial separation between the macro BSs and SBSs as well as the clustering nature of the SBSs~\cite{AfsDhiC2015a}.   
From analysis perspective, it is important to extend the results to more general channel models, such as $\kappa-\mu$ shadowed fading
channels~\cite{chun2016comprehensive}, and correlated shadowing.   From  application perspective, the results can be extended to the analysis of cache enabled network to study  metrics like total hit probability and caching throughput; see \cite{afshang2016optimal1,bacstuug2016edge}. Finally,  this framework  can be extended to the analysis of other key performance metrics such as ergodic spectral efficiency~\cite{ErgodicSELozano2017} and bit error rate.


\appendix
\subsection{Proof of Lemma~\ref{lem: PDFs of intra-cluster distances}}
\label{proof: PDFs of intra-cluster distances}

Let us denote the location of SBS chosen uniformly at random in the representative cluster by ${\bf z}_0={\bf x}_{0}+{\bf y}_{\rm s} \in \R^2$, where  ${\bf z}_0=(z_1,z_2)$ and ${\bf x}_0= (\nu_0,0)$. The conditional CDF of distance $U$ with realization $u=\sqrt{z_1^2+z_2^2}\in \R_{+}$  is: 
\begin{align*}
F_{U}(u|{\bf x}_0)=F_{U}(u|\nu_0)&=\int \int_{\sqrt{z_1^2+z_2^2} \le u} f_{{\bf Y}_{\rm s}}\big( z_1-\nu_0, z_2\big) {\rm d}z_2 {\rm d} z_1\\
&= \int_{z_1=-u}^{z_1=u} \int_{z_2=-\sqrt{u^2- z_1^2}}^{z_2=\sqrt{u^2- z_1^2}}  f_{{\bf Y}_{\rm s}}\big( z_1-\nu_0, z_2\big)  {\rm d}z_2 {\rm d} z_1,
\end{align*}
where the PDF of $U$ is obtained by using Leibniz's rule for differentiation~\cite{samko1993fractional}. Now recall that the typical user is located at the origin, and users are distributed around cluster center with PDF $f_{{\bf Y}_{\rm u}}({\bf y_u})$. Thus,  the  relative location of the
cluster center with respect to the typical user, i.e ${\rm x}_0$,  has the same distribution as that of ${\bf Y}_{\rm u}$.  The PDF of $V_0=\|{\bf Y}_u\|$ can be derived by using the same argument applied in the derivation of $f_{U}(.)$.

%
%
%

\subsection{Proof of Lemma~\ref{lem : association prob}}
\label{proof : association prob}
The conditional  association probability to the macro-tier for a given value of $\nu_0$ is: 
\begin{align}\notag
{\cal A}_{\rm m}^{\rm P1}(\nu_0)&=\E_{R_{\rm m}}[{\bf1}\{P_{\rm m}  R_{\rm m}^{-\alpha}>P_{\rm s}  R_{\rm s}^{-\alpha}\}|\nu_0]
=\int_0^{\infty}\P \Big[ R_{\rm s} >\big(\frac{P_{\rm s} }{P_{\rm m}  }\big)^{1/\alpha}r_{\rm m}|\nu_0 \Big] f_{R_{\rm m}}(r_{\rm m}) {\rm d} r_{\rm m},\\\notag
&=\int_0^{\infty} \Big[1-F_{R_{\rm s} }\big(\big(\frac{P_{\rm s} }{P_{\rm m} }\big)^{1/\alpha}r_{\rm m}|\nu_0\big) \Big] f_{R_{\rm m}}(r_{\rm m}) {\rm d} r_{\rm m},
\end{align}
using which the  association probability  to the small cell  tier is ${\cal A}^{\rm P1}_{\rm s} (\nu_0)=1-{\cal A}^{\rm P1}_{\rm m}(\nu_0).$
\subsection{Proof of Lemma~\ref{lem: serving distance}}
\label{Proof: serving distance}
For a given typical user located at distance $\nu_0$ from its cluster center,  the event $X_{\rm m}^{\rm P1}> x_{\rm m}$ is equivalent to that of $R_{\rm m}> x_{\rm m}$ when the typical user connects to the macro BS, i.e.,  event $S_{{\rm m}}^{\rm P1}$. Thus, the conditional  CCDF of $X_{\rm m}^{\rm P1}$ can be derived as:
\begin{align}
\P({X_{\rm m}^{\rm P1}>x_{\rm m}}|\nu_0)&=\P(R_{\rm m}> x_{\rm m}|S_{{\rm m}}^{\rm P1},\nu_0)=\frac{\P[R_{\rm m}> x_{\rm m},S_{{\rm m}}^{\rm P1}|\nu_0]} {\P[S_{{\rm m}}^{\rm P1}|\nu_0]} \\
&=\frac{1} {{\cal A}_{\rm m}^{\rm P1}(\nu_0)}  \int_{x_{\rm m}}^{\infty}\P(R_{\rm s}> \xi_{\rm s m} \: r_{\rm m}|\nu_0) f_{R_{\rm m}}(r_{\rm m}) {\rm d} {r_{\rm m}},
\end{align}
and hence the PDF of $X_{\rm m}^{\rm P1}$ is:
\begin{align}
f_{X_{\rm m}^{\rm P1}}(x_{\rm m}|\nu_0)=\frac{{\rm d}}{{\rm d} x_{\rm m}} (1-\P(X_{\rm m}^{\rm P1}> x_{\rm m}|\nu_0))=\frac{1} {{\cal A}_{\rm m}^{\rm P1}(\nu_0)}(1-F_{R_{\rm s}}(\xi_{\rm s m} x_{\rm m}|\nu_0)) f_{R_{\rm m}}(x_{\rm m}).
\end{align}
The derivation of $f_{X_{\rm s}^{\rm P1}}(\cdot|\nu_0)$  follows on the same lines  as that of $f_{X_{\rm m}^{\rm P1}}(\cdot|\nu_0)$, and is hence skipped.

\subsection{Proof of Lemma~\ref{Lem: Interfering distances from intra}}
\label{Proof: Interfering distances from intra}
 Denote by $\{U_j\}_{j=1}^{n_{{\rm s}_0}}$ the sequence of distances from open access SBSs to the typical user.
From  association Policy~1,  it can be deduced that there {are} no open-access BSs {within} distance $\xi_{\rm s m} x_{\rm m}$ of the typical user when this user is served by the macro BS. Thus, the sequence of distances from interfering  open-access SBSs to the typical user can be defined as  ${\cal F}^{\rm P1}_{\rm m}\equiv\{W^{\rm P1}_{{\rm m},j}: W^{\rm P1}_{{\rm m},j}=U_j, U_j>\xi_{\rm s m} x_{\rm m}\}$.  The conditional  joint density function of  the elements in  ${\cal F}^{\rm P1}_{\rm m}$ is:
\begin{align*}
&\P(W^{\rm P1}_{{\rm m},1}<w_{{\rm m},1},... , W^{\rm P1}_{{\rm m},n_{{\rm s}_0}}<w_{{\rm m},n_{{\rm s}_0}}| V_0=\nu_0)\\
=&\P(U_1<w_{{\rm m},1},... , U_{n_{{\rm s}_0}}<w_{{\rm m},n_{{\rm s}_0}}| U_1>\xi_{\rm s m} x_{\rm m},..., U_{n_{\rm  s}}>\xi_{\rm s m} x_{\rm m} ,V_0=\nu_0) \\
=&\left\{
 \begin{array}{cc}
\frac{\P(U_1<w_{{\rm m},1},... , U_{n_{{\rm s}_0}}<w_{{\rm m},n_{{\rm s}_0}}| V_0=\nu_0)}{\P(U_1>\xi_{\rm s m} x_{\rm m},..., U_{n_{\rm  s}}>\xi_{\rm s m} x_{\rm m} |V_0=\nu_0) }, & w_{{\rm m},j}\geq\xi_{\rm s m} x_{\rm m}\\
 0, &0\leq w_{{\rm m},j}< \xi_{\rm s m} x_{\rm m}\
 \end{array}\right., \quad j=\{1,2,..., n_{{\rm s}_0}\}\\
\stackrel{(a)}{=}&\left\{
 \begin{array}{cc}
\prod_{j=1}^{n_{{\rm s}_0}} \frac{F_{U_j} (w_{{\rm m},j}|\nu_0)}{1-F_{U_j}(\xi_{\rm s m}x_{\rm m}|\nu_0)}, & w_{{\rm m},j}\geq\xi_{\rm s m} x_{\rm m}\\
 0, &0\leq w_{{\rm m},j}< \xi_{\rm s m} x_{\rm m}\
 \end{array}\right., \quad j=\{1,2,..., n_{{\rm s}_0}\},
\end{align*}
where $(a)$ follows from  the fact that the elements in $\{U_j\}_{j=1}^{n_{{\rm s}_0}}$ are conditionally i.i.d. (see Lemma~\ref{lem: PDFs of intra-cluster distances}).  The product of the same functional  form in  the joint  CDF implies that  the elements in ${\cal F}^{\rm P1}_{\rm m}$ are conditionally i.i.d. with CDF  $ F_{W^{\rm P1}_{{\rm m},j}}(w_{{\rm m},j}|\nu_0, x_{\rm m} )=\frac{F_{U_j} (w_{{\rm m},j}|\nu_0)}{1-F_{U_j}(\xi_{\rm s m}x_{\rm m}|\nu_0)}$.  Using this result,  the  PDF of $W^{\rm P1}_{{\rm m},j}$   can be  obtained  by taking derivative of  $F_{W^{\rm P1}_{{\rm m},j}}(w_{{\rm m},j}|\nu_0,x_{\rm m})$ with respect to $w_{{\rm m},j}$. In the final result, index $j$ is dropped for notational simplicity. The derivation of $f_{W^{\rm P1}_{{\rm s}}}(\cdot|\nu_0,x_{\rm s})$ follows on the same lines as that of ${W^{\rm P1}_{{\rm m}}}$, and is hence skipped.

\subsection{Proof of Lemma \ref{lem: Laplace intra}}
\label{Proof: Laplace intra}
The Laplace transform of intra-cluster interference distribution at a typical user served by macrocell  conditioned on $V_0$ and $X^{\rm P1}_{\rm m}$ is:
\begin{align*}
{\cal L}_{{\cal I}_{\rm sm}^{\rm intra}}(s|\nu_0,x_{\rm m})\stackrel{(a)}{=} \E\Big[\exp\Big(-s\sum_{{\bf y}_{\rm s}\in {\cal B}_{\rm s}^{{\bf x}_0}} P_{\rm s} h_{\rm s} \|{\bf x}_0+{\bf y}_{\rm s}\|^{- \alpha}\Big)\Big]
\stackrel{(b)}{=}\E\Big[\prod_{{\bf y}_{\rm s}\in {\cal B}_{\rm s}^{{\bf x}_0}} \frac{1}{1+s P_{\rm s}  \|{\bf x}_0+{\bf y}_{\rm s}\|^{- \alpha} }\Big],
\end{align*}
where $(a)$ follows from the definition
of Laplace transform and $(b)$  follows from the expectation over $h_{\rm s}\sim \exp(1)$. 
The final result follows from the change of variable $\|{\bf x}_0+{\bf y}_{\rm s}\| \rightarrow w_{\rm m}$, and converting  from
Cartesian to polar coordinates,  followed by the fact that  the elements of $\{W_{\rm m}\}$ are conditionally i.i.d., with PDF $f_{W^{\rm P1}_{\rm m}}(w_{\rm m}|\nu_0,x_{\rm m})$ given by Lemma~\ref{Lem: Interfering distances from intra}, followed by  expectation over the number of simultaneously active  SBSs within the representative cluster with PDF
 \begin{align}
\P(|{\cal B}_{\rm s}^{{\bf x}_0}|=\ell)=\frac{  {\bar{n}_{\rm as}}^{\ell} e^{-\bar{n}_{\rm as}}}{\ell! \sum_{k=0}^{n_{\rm s_0}} \frac{{\bar{n}_{\rm as}}^{k} e^{-\bar{n}_{\rm as}}}{k!}},\quad 0\le \ell \le n_{{\rm s_0}},
\end{align}
where $|{\cal B}_{\rm s}^{{\bf x}_0}|$ is  Poisson distributed conditioned on the total being less than $n_{\rm s_0}$.
The derivation of  ${\cal L}_{{\cal I}_{\rm ss}^{\rm intra}}(.|\nu_0,x_{\rm s})$ follows on the same lines as that of ${\cal L}_{{\cal I}_{\rm sm}^{\rm intra}}(.|\nu_0,x_{\rm m})$, where the serving SBS is removed from the set of possible interfering SBSs.
The PDF of number  of simultaneously active  SBSs within the representative cluster $|{\cal B}_{\rm s}^{{\bf x}_0}|$ conditioned on having at least one active SBS (serving SBS)  is:
 \begin{align}
\P(|{\cal B}_{\rm s}^{{\bf x}_0}|=\ell)=\frac{  {\bar{n}_{\rm as}}^{\ell-1} e^{-\bar{n}_{\rm as}}}{(\ell-1)! \sum_{k=1}^{n_{\rm s_0}} \frac{{\bar{n}_{\rm as}}^{k-1} e^{-\bar{n}_{\rm as}}}{(k-1)!}},\quad 1\le \ell \le n_{{\rm s_0}},
\end{align}
which is  truncated weighted Poisson  distribution.

\subsection{ Proof of Lemma \ref{lem: Laplace inter}}
\label{Proof: Laplace inter}
Note that the Laplace transform of inter-cluster interference does not depend on the choice of the serving BS. Denoting by $j \in \{{\rm s}, {\rm m}\}$ the index of  the serving BS, the Laplace transform of  inter-cluster interference ${\cal I}^{\rm inter}_{{\rm s}j}$ is ${\cal L}_{{\cal I}^{\rm inter}_{{\rm s}j}}(s)$
\begin{align*}
&\stackrel{(a)}{=}\E\Big[\exp\Big(-s \sum_{{\bf x} \in \Psi_{\rm p} \setminus {\bf x}_0} \sum_{{\bf y}_{\rm s}\in {\cal B}_{\rm s}^{{\bf x}}} P_{\rm s} h_{\rm s} \|{\bf x}+{\bf y}_{\rm s}\|^{- \alpha} \Big)\Big]\\
 &\stackrel{(b)}{=}\E_{\Psi_{\rm p}} \prod_{{\bf x} \in \Psi_{\rm p}\setminus {\bf x}_0} \E_{ {\cal B}_{\rm s}^{{\bf x}}} \prod_{ {\bf y}_{\rm s} \in {\cal B}_{\rm s}^{{\bf x}}}  \E_{h_{\rm s}} \exp\Big(-s P_{\rm s} h_{\rm s} \|{\bf x}+{\bf y}_{\rm s}\|^{- \alpha} \Big)\\
&\stackrel{(c)}{=}\E_{\Psi_{\rm p}} \prod_{{\bf x} \in \Psi_{\rm p}\setminus {\bf x}_0} \E_{ {\cal B}_{\rm s}^{{\bf x}}} \prod_{ {\bf y}_{\rm s} \in {\cal B}_{\rm s}^{{\bf x}}} \frac{1}{ 1+s   P_{\rm s} \|{\bf x}+{\bf y}_{\rm s}\|^{- \alpha} } \\&\stackrel{(d)}{=}\E_{\Psi_{\rm p}} \prod_{{\bf x} \in \Psi_{\rm p}\setminus {\bf x}_0} \exp\Big(-\bar{n}_{
\rm a s} \int_0^{\infty}  \frac{s P_{\rm s} t_{\rm s}^{- \alpha}}{ 1+s P_{\rm s} t_{\rm s}^{- \alpha} } f_{T_{\rm s}}( t_{\rm s}|\nu) {\rm d} {t_{\rm s}}\Big)\\
&\stackrel{(e)}{=}\exp\Big(-2 \pi \lambda_{\rm p} \int_0^{\infty} \Big(1-  \exp\Big(-\bar{n}_{
\rm a s} \int_0^{\infty}  \frac{s P_{\rm s} t_{\rm s}^{- \alpha}}{ 1+s P_{\rm s} t_{\rm s}^{- \alpha} } f_{T_{\rm s}}( t_{\rm s}|\nu) {\rm d} {t_{\rm s}}\Big)\Big) \nu {\rm d} {\nu}\Big),
\end{align*}
where $(a)$ follows from definition of Laplace transform, $(b)$ follow from the assumption that  fading gains
across all interfering links are independent, $(c)$ follows from the expectation over $h_{\rm s}\sim \exp(1)$, $(d)$ follows from the change of variable $\|{\bf x}_0+{\bf y}_{\rm s}\| \rightarrow t_{\rm s}$, and converting  from Cartesian to polar coordinates,  followed by the  fact that number of points per cluster are Poisson distributed, and $(e)$ follows from probability generating functional (PGFL) of PPP.

{ \setstretch{1.3}
\bibliographystyle{IEEEtran}
\bibliography{Arxiv_J4_fixed_ClusteredHetNEt_V7.bbl}
}
\end{document}